\documentclass[prb,twocolumn,showpacs,preprintnumbers,amsmath,amssymb]{revtex4}
\usepackage{graphicx}
\usepackage{color}

\newcommand{\vc}{\vec}

\newcommand{\EF}{\ensuremath{E_{\mathrm{F}}}}

\begin{document}

\title{Lifetime reduction of surface states at Cu, Ag and Au(111) caused by impurity scattering}

\author{Swantje Heers}
\author{Phivos Mavropoulos}\email{Ph.Mavropoulos@fz-juelich.de} 
\author{Samir Lounis}
\author{Rudolf Zeller}
\author{Stefan Bl\"ugel}

\affiliation{Peter Gr\"unberg Institut and Institute for Advanced Simulation, Forschungszentrum
  J\"ulich and JARA, D-52425 J\"ulich, Germany}

\begin{abstract}
  We present density-functional results on the lifetime of the (111)
  surface state of the noble metals. We consider scattering on the
  Fermi surface caused by impurity atoms belonging to the $3d$ and
  $4sp$ series. The results are analyzed with respect to film
  thickness and with respect to separation of scattering into bulk or
  into surface states. While for impurities in the surface layer the
  overall trends are similar to the long-known bulk-state scattering,
  for adatom-induced scattering we find a surprising behavior with
  respect to the adatom atomic number.  A plateau emerges in the
  scattering rate of the $3d$ adatoms, instead of a peak
  characteristic of the $d$ resonance. Additionally, the scattering
  rate of $4sp$ adatoms changes in a zig-zag pattern, contrary to a
  smooth parabolic increase following Linde's rule that is observed in
  bulk. We interpret these results in terms of the weaker
  charge-screening and of interference effects induced by the lowering
  of symmetry at the surface.
\end{abstract}



\maketitle

\section{Introduction}

The lifetime of the noble-metal (111) surface states has been the
subject of numerous theoretical and experimental investigations. These
states are formed due to the breaking of translational symmetry at the
surface in combination with the band gap existing in the
surface-projected bandstructure in the [111] direction. Due to their
localized nature, they play a role in a variety of surface-related
phenomena, such as catalysis, adsorbate interactions, or surface
transport, and can be probed by spectroscopic surface-sensitive
techniques such as scanning tunneling microscopy and spectroscopy (STM
and STS), photoemission, etc..\cite{Kroeger_STM_surface_states} They
furthermore constitute the fundament for the appearance of
quantum-interference phenomena such as the famous quantum mirage in
quantum-coral systems.\cite{Manoharan00}

Although surface states are orthogonal to bulk states when only the
periodic crystal and surface potential is considered, in reality there
are always coupling mechanisms of the surface states to each other and
to the bulk states. Mainly, these mechanisms are electron-electron
interactions (including interactions with collective electron modes
such as plasmons), henceforth labeled ``ee,'' electron-phonon
interactions, henceforth labeled ``ep,'' but also scattering at
defects of the surface such as step edges, islands or adsorbate atoms,
henceforth labeled ``ed.'' The total lifetime $\tau$ associated with
these mechanisms is related to a sum of the scattering rates by
\begin{equation}
\frac{1}{\tau} = \frac{1}{\tau_{\rm ee}} + \frac{1}{\tau_{\rm ep}} +
\frac{1}{\tau_{\rm ed}}\;\mbox{.}
\label{eq:1}
\end{equation}
At low temperature, the phonon contribution becomes negligible, while
the contribution of the electron-electron interaction depends on the
electron energy $E$ with respect to the Fermi level
\EF.\cite{Vitali2003L47} For $E$ close to \EF, and assuming
Fermi-liquid behavior, the contribution to the decrease of the
lifetime from electron-electron interactions behaves as $\tau_{\rm
  ee}^{-1}\sim (E-\EF)^2$ and is thus also ultimately frozen
out. However, the scattering at defects remains finite.

The lifetime $\tau$ of surface states is an experimentally measurable
quantity. Most commonly used methods for this purpose are angular
resolved photoemission spectroscopy
(ARPES)\cite{PhysRevB.51.13891,PhysRevB.63.115415,Reinert2004229} and
STS.\cite{progr_surfa_science_Kroeger,Kroeger_STM_surface_states}
ARPES can detect the full surface band structure, $E_{\vec{k}}$ and
$\tau_{\vec{k}}$ (with $1/\tau_{\vec{k}}$ appearing as a broadening
$\Gamma$ of the electron bands), but defect scattering is necessarily
averaged over a large surface area and cannot be disposed of. In fact,
one of the first experiments where the importance of defect-scattering
was detected was based on ARPES.\cite{PhysRevLett.50.526} It was shown
that the lifetime $\tau$ of the Cu(111) surface state actually
\emph{decreased} for energies close to \EF, contrary to the
predictions of Fermi-liquid theory; this effect was attributed to
strong defect-scattering at those energies.

STS, on the other hand, follows the onset of the surface state in the
$dI/dV$ signal which is proportional to the density of states,
changing shape from step-like (as it would be for a two-dimensional
electron system), to obtaining a certain width $\Gamma \sim \tau^{-1}$
due to scattering.\cite{PhysRevLett.94.036805,PhysRevLett.81.4464}
Additionally, $\tau$ can be found via the phase coherence length
$L_{\phi}$ by probing a quantum interference pattern in the $dI/dV$
curves produced by nanoscale
resonators.\cite{1367-2630-3-1-322,PhysRevB.71.155417,Li199995} In
particular, STS has the advantage that one can choose a defect-free
surface region, in order to isolate the electron-electron and
electron-phonon scattering from the
defect-scattering.\cite{PhysRevLett.81.4464,Kliewer26052000,footnote1}

Whereas in most of the experiments localized defects have been avoided
as much as possible, there are a few experimental studies which
followed the idea of controlling the defect concentration, in order to
extrapolate to a defect-free surface. The first of these experiments
has been performed with ARPES for different concentration of potassium
atoms on Cu (111),\cite{PhysRevB.33.4364} where the structural
disorder was investigated and quantified with LEED. In references
\onlinecite{PhysRevB.56.3632,Theilmann199933,PhysRevB.61.16168},
similar experiments have been reported, in which the surface-state
lifetimes at several coverages of Cu adatoms on Cu (111) have been
measured in order to extrapolate to zero-defect concentration.

Theoretical approaches to the surface-state lifetime address, in most
cases, the electron-electron and electron-phonon interaction, under
the assumption of an ideally defect-free surface. Progress has been
made in calculation of the effects of the electron-electron
interaction, as the initial degenerate electron gas
model\cite{PhysRev.126.1453} was replaced by the more elaborate $GW$
approximation\cite{GWA} of many-body theory. Applied to image
potential states,\cite{PhysRevLett.80.4947} this approach led to a
nice agreement with experiment,\cite{Kliewer26052000} accounting
for both intra-band and inter-band scattering. Concerning
electron-phonon scattering, already calculations based on the simple
Debye model lead to a satisfying agreement between numerical and
experimental data,\cite{Kliewer26052000} while more accurate
$ab~initio$ calculations of the phonon density of states comprise the
investigation of bulk and surface
phonons.\cite{PhysRevLett.88.066805,PhysRevB.67.235423} Generally, at
low temperatures the electron-phonon contribution of the scattering
rate behaves as $\Gamma = 2 \pi \lambda k_B T$ with $\lambda$
being the electron-phonon coupling parameter. This linear behavior is
also found by experiment.\cite{PhysRevB.51.13891,Reinert2004229}

However, in particular on scattering of surface states at adsorbate
atoms, only few calculations have been published to date, mainly based
on the wavepacket propagation method, which is a real-space, real-time
approach.\cite{PhysRevB.78.245410} A systematic theoretical \emph{ab
  initio} investigation of lifetime reduction caused by (elastic)
scattering at different species of adatoms or substitutional
impurities in the (111) surfaces of the noble metals is still
lacking. Bridging this gap is the purpose of this paper. Furthermore,
many of the aforementioned lifetime measurements (especially in STS
experiments) have been performed at the lower band edge of the
surface-state band, where the contribution to the surface-state
linewidth caused by defect scattering merges with the
electron-electron and the electron-phonon contribution. This is
different at the Fermi energy (at least at low temperatures), where
defect scattering becomes the dominant mechanism.  In addition, the
Fermi level is the relevant energy for transport processes. These are
the main reasons, why we have performed our calculations at the Fermi
level.

The paper is organized as follows: We start with an introduction to
the theoretical formalism, i.e.~multiple scattering theory, based on
the Korringa-Kohn-Rostoker (KKR) Green function method. We proceed
with an analysis of computational and numerical aspects of the
calculations, demonstrating differences between full potential
calculations and those performed within the atomic sphere
approximation.  We study the lifetime reduction due to scattering at
$3d$ and $4sp$ impurities in the first surface layer and in the fcc
adatom position on the noble metal thin films Cu, Ag and Au, which
turn out to show similar trends. While the trend for the scattering
rates for impurities in the first surface layer is found to be similar
to that calculated in bulk materials\cite{Graefenstein88} (known also
from residual resistivity results\cite{Mertig82}), the trend for
scattering at adatoms remarkably differ. Therefore, we focus on
adatoms, and analyze the trend of scattering rates for $3d$ and $4sp$
adatoms on Cu(111). This analysis includes an angular-momentum
resolved study of surface-state lifetimes as well as a resolution of
scattering into bulk and surface states.  Finally, the trend observed
for magnetic impurities and adatoms is investigated.

A note should be made here on our approximation concerning magnetic
impurities. Below the Kondo temperature, the magnetic moment of these
defects is quenched by many-body effects.\cite{hewson} The Kondo
temperature varies over orders of magnitude with respect to impurity
type and host, depending exponentially on the position of the $d$
resonance and on the hybridization. Therefore, there is no 'unique'
temperature above which all $3d$ impurities become simultaneously
magnetic. At a sufficiently low temperature, however, they should all
be non-magnetic in the sense of the Kondo effect. In this case, the
non-magnetic density of states in the local-density approximation does
not represent the physical density of states, except exactly at the
Fermi energy \EF, where it is probably a good approximation.

This follows from considerations on the Anderson impurity model. In
the case of a single localized impurity orbital, the exact solution
gives the same phase shift and density of states at \EF\ as the
restricted mean-field solution (i.e., the mean-field solution where
the two spin populations are constrained to be equal to each other,
which is basically equivalent to a non-magnetic calculation). In a
more general case the Friedel sum rule reads
\begin{equation}
\Delta Z = -\frac{1}{\pi}\sum_{lm\sigma}\delta_{lm\sigma}(\EF)
\end{equation}
where $\Delta Z$ is the valence difference between impurity and host
while $\delta_{lm\sigma}(\EF)$ is the phase shift at \EF\ for angular
momentum $lm$ and spin $\sigma$ and is directly related to the
scattering matrix and scattering rate. If one assumes that only the
$d$ electrons screen the impurity charge of a transition-metal
impurity and that all $d$ orbitals are equivalent and have the same
phase shift $\delta_{d}(\EF)$, and additionally takes into account
that $\delta_{d}(\EF)$ is spin-independent in the Kondo regime, then
the Friedel sum rule reads $\Delta Z=-(10/\pi)\delta_{d}(\EF)$. Thus
the phase shift at \EF\ is fixed to the same value whether one
considers the restricted-mean-field or the exact solution. In reality
the assumption that only the $d$ electrons screen the impurity is an
approximation, as is the assumption that the phase shift of all $d$
states is equivalent, since there is a small crystal-field
spitting. In practice, however, the two assumptions are relatively
well fulfilled, as one can see from comparison of experiment with
calculations on the residual resistivity of simple-metal dilute alloys
with $3d$ impurities, where the impurity electronic structure was
calculated within the density-functional theory and restricted to be
non-magnetic (see, e.g.,
Refs.~\onlinecite{Mertig82,Papanikolaou94}). Therefore, the results
obtained here by the non-magnetic calculations should be considered to
be a reasonable approximation, since only Fermi-level phase shifts
enter the calculation.  The approximation clearly breaks down for
properties that probe the phase shift off the Fermi level, e.g. in the
thermopower where the derivative $d\delta_l(E)/dE$ at \EF\ enters. In
this case the density-functional calculations\cite{Mavropoulos95}
cannot account for the giant thermopower encountered for the Kondo
impurities Mn and Fe in Al.

\section{Theory}
\label{sec:theory}

Details of the KKR method for electronic structure calculations and
impurity scattering have been published
elsewhere.\cite{Mertig99,Papanikolaou02,Ebert11} Here we
focus on the definition of some quantities which will be useful in
the discussion.

In the KKR method, the Bloch wavefunction $\psi_{\vec k}(\vec r+\vec
R^n)$ at a point $\vec r$ within an atomic cell $n$ sited at $\vec
R^n$ is expanded as
\begin{equation}
\psi_{\vec k}(\vec r+\vec R^n;E)=\sum_L
c_{\vec{k}\, nL}(E)  R_{nL}(\vec r;E) \; \mbox{,}
    \label{eqn:def_psi}
\end{equation}
where the $R_{nL}(\vec r;E)$ are regular scattering solutions of the
Schr\"odinger equation of the atom $n$ embedded in free space with
boundary condition of an incoming spherical wave of angular momentum
$L=(l,m)$, while $E$ denotes the energy eigenvalue. The coefficients
$c_{\vec{k}\, nL}(E)$ are found by the solution of the KKR secular
equation for lattice site 0 and propagated by a Bloch factor to any
other site.

In the presence of an impurity, an incoming Bloch wave $\psi_{\vec k}$
evolves after scattering into an outgoing wavefunction $\psi_{\vec
  k}^{\rm imp}$ (with $\vec{k}$ here denoting the initial incoming
state). The associated transition amplitude ($T$ matrix),
\begin{equation}
\label{eqn:def_T_kk'}
 T_{\vec k \vec k'}(E):= \int d^3 r \; \psi_{\vec k'}^{\star} (\vec r;E) \Delta V(\vec r) \psi_{\vec k}^{\mathrm{imp}} (\vec r ;E),
\end{equation}
is calculated using the expression
\begin{equation}
    T_{\vec k \vec k'}(E) = \sum_{n n'} \sum_{LL'} c^{\star}_{\vec k nL}(E)
    T_{LL'}^{nn'}(E) c_{\vec k' n'L'}(E)
    \label{eqn:T_kk_T_nn}
\end{equation}
where the algebraic form of the $T$ matrix
\begin{equation}
    T_{LL'}^{nn'} := \sum_{L''} \Delta^n_{LL''} \left( \delta_{L''L'}
    \delta_{nn'} + \sum_{L'''} G_{L'' L'''}^{{\rm imp}, nn'} \Delta t_{L'''L'}^{n'} \right) 
    \label{eqn:def_T_LLnn}
\end{equation}
contains information on the impurity structural Green function $G_{L
  L'}^{{\rm imp}, nn'}(E)$ and on the difference in the atomic scattering
properties between the embedded-impurity and substituted-host atom at
site $n$:
\begin{eqnarray}
    \Delta t_{LL'}^n(E):=\int_{\rm{cell}\;n} d^3 r \; R_{nL} (\vec r;E)
    \Delta V_n (\vec r) R_{nL'}^{\rm{imp}} (\vec r;E)        
\label{eqn:def_delta_mat1}    \\
 \Delta_{LL'}^n(E):=\int_{\rm{cell}\;n} d^3 r \; R_{nL}^{\star} (\vec r;E)
    \Delta V_n (\vec r) R_{nL'}^{\rm{imp}} (\vec r;E)\; \mbox{,}
    \label{eqn:def_delta_mat}
\end{eqnarray}
with $R_{nL'}^{\rm{imp}} (\vec r;E)$ being the regular scattering
solutions of the impurity atom in free space in analogy to the
host-atom solutions $R_{nL} (\vec r;E)$ (the r.h.s of
Eqs.~(\ref{eqn:def_delta_mat1}) and (\ref{eqn:def_delta_mat}) differ
only by the complex conjugation of $R_{nL} (\vec r;E)$).  Since
$\Delta t_{LL'}^n(E)$ and $\Delta_{LL'}^n(E)$ vanish at sites where
the perturbation of the potential, $ \Delta V_n (\vec r)$, is zero,
the summation in Eq.~(\ref{eqn:def_T_LLnn}) is limited to the sites
for which the self-consistent potential is appreciably affected by the
presence of the impurity; usually the impurity site and its nearest
neighbors are sufficient.  Only these sites are taken into account
also for the calculation of the impurity structural Green function via
an algebraic Dyson equation
\begin{eqnarray}
G_{L L'}^{{\rm imp}, nn'} &=& G_{L L'}^{{\rm host}, nn'}  \nonumber \\
&+& \sum_{n''L''L'''} 
G_{L L''}^{{\rm host}, nn''}\Delta t^{n''}_{L''L'''}G_{L''' L'}^{{\rm imp}, n''n'}
\end{eqnarray}
from the host Green function $G_{L L'}^{{\rm host}, nn'}(E)$

Finally, the probability for an electron
scattering from a state $\vec k$ to a state $\vec k'$ is given by  
\begin{equation}
 P_{\vec k \vec k'}=\frac{2 \pi}{\hbar} Nc \left| T_{\vec k \vec k'} \right|^2
 \delta (E_{\vec k}-E_{\vec k'})\; \mbox{,}
 \label{eqn:trans_pr}
\end{equation}
where $N$ is the total number of atoms in the crystal and $c$ is the
impurity concentration ($Nc$ is the number of randomly positioned
impurities). Assuming that each impurity scatters independently, the
lifetime of a state $\vec k$ can be found by summing up the
probabilities for scattering of a state $\vec k$ into all possible
final states $\vec k'$
\begin{equation}
 \tau_{\vec k}^{-1}=\sum_{\vec k'} P_{\vec k \vec k'}= \frac{2 \pi}{\hbar} Nc
 \sum_{\vec k'} \left| T_{\vec k \vec k'} \right|^2 \delta (E_{\vec k}-E_{\vec
 k'}) \; \mbox{.}
    \label{eqn:inv_lifetime}
\end{equation}
The summation over $\vec k'$ can be transformed to an integral, such that the
inverse lifetime is obtained by integration over the Fermi surface $S(E_\mathrm{F})$
\begin{equation}
\tau_{\vec k}^{-1} 
= \frac{1}{V_{\mathrm{BZ}}} \frac{2 \pi N^2 c}{ \hbar^2}\int_{S(E_\mathrm{F})}
 \frac{dS_{\vec k'}}{v_{\vec k'}} \; \left| T_{\vec k \vec k'} \right|^2  \; \mbox{,}
\label{eqn:inv_lifetime_int}
\end{equation}
where $dS_{\vec k'}$ is the Fermi surface element, $v_{\vec k}$ is the
Fermi velocity and $V_{\rm BZ}$ the Brillouin zone volume.

\section{Computational and numerical aspects of the calculations}

The electronic structure of the investigated finite-thickness films
and defects has been calculated self-consistently within
density-functional theory using the principal layer technique
implemented in the KKR Green function
method\cite{PhysRevB.55.10074,PhysRevB.52.8807} using the experimental
lattice parameter and an angular momentum cutoff of
$l_{\mathrm{max}}=3$. Exchange and correlation effects were included
within the local-density approximation to density-functional theory in
the parametrization of Vosko et al,\cite{Vosko_Wilk_Nusair} while
relativistic effects were taken into account in the scalar
relativistic approximation (ignoring spin-orbit coupling). Except from
section \ref{sec:comp_ASA_FP}, where the difference between full
potential (FP) calculations and those performed in the atomic sphere
approximation (ASA) is compared, the ASA was used. The perturbed
region, where charge relaxation is allowed, is restricted in our
calculations to a cluster of 13 sites, thus the shell of nearest
neighbors, unless otherwise stated. Structural relaxations are not
taken into account, as they could have a quantitative but not
qualitative effect to the trends that are related, as we find, to
symmetry reduction, reduced screening of adatoms, and the variation of
localization of the surface states as a function of film thickness.

Our calculations are performed on finite-thickness films. The
situation of a half-infinite crystal with a single surface is
approximated by increasing the film thickness up to 40
layers. Furthermore, the vacuum region is described by empty atomic
sites at 3 (or in the FP calculations 4) layers above the surface;
these are called `vacuum layers', composed of `vacuum sites'.

\subsection{Fermi surfaces of copper, silver and gold (111) films}

\begin{figure}[h]
  \begin{center}
       \includegraphics*[width=0.2\textwidth,angle=-90]
         {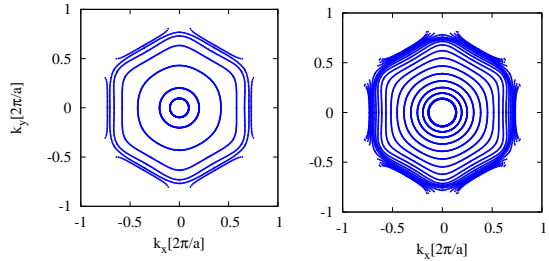}
         \caption{Fermi surfaces of Cu(111) films direction for 6
           (\emph{left}) and 18 layers (\emph{right}) shown within the
           first Brillouin zone.  In the calculations three `vacuum
           layers' have been added on each side. The splitting of the
           two surface states (represented by the two innermost rings)
           decreases with the number of layers such that it is not
           visible in this scale for the 18 layers-film (the two
           innermost rings fall practically on top of each other and
           appear as a thicker circle).}
  \label{fig:FS_Cu}
  \end{center}
\end{figure}

The three investigated materials, copper, silver and gold crystalize
in the fcc structure, have a very similar electronic structure and
therefore their Fermi surfaces are very similar to each other, too. As
these systems are characterized by a two-dimensional periodicity,
their Fermi surface consists of one-dimensional curves that form
ring-like structures, except in the vicinity of the Brillouin-zone
boundary, where hexagonal-like structures occur. The number of rings
on the Fermi surface scales linearly with the number of layers, as can
be seen in Fig.~\ref{fig:FS_Cu} for the example of a copper (111)
Fermi surface for films with 6 (left) and 18 copper layers
(right). The two innermost rings represent the surface bands which are
formed in the gap of the surface-projected
band-structure.\cite{PhysRev.56.317}

\begin{figure}[]
  \centering 
       \includegraphics*[width=0.34\textwidth, angle=-90]
         {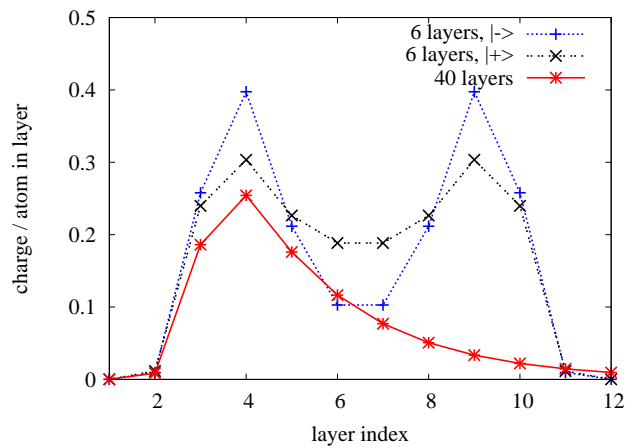}
         \caption{Charge (in units of the electron charge) per atom in
           the different layers for the two surface states of a copper
           film with six layers and for the innermost Fermi ring of a
           copper film with 40 layers (the three outmost layers
           correspond to vacuum). Here, $|+\rangle$ and $|-\rangle$
           denote the bonding and antibonding surface state, i.e.,
           corresponding to the innermost/second innermost Fermi ring,
           respectively. For the thinner film, the charge in each
           layer is larger than for the 40 layer-film, because the
           number of layers in which the surface states can penetrate
           is limited. The two states of the 40 layer film are almost
           degenerate and nearly indistinguishable in their
           charge-per-layer picture.}
  \label{fig:Local_SF_states}
\end{figure}

Surface states are localized at the atomic layers close to the surface
and decay exponentially into the bulk and into the vacuum. Since in a
finite film there are two surfaces (say ``Left'' and ``Right''), also
two surface states $\psi_{\rm L}$ and $\psi_{\rm R}$ appear and
interact, forming a bonding and an anti-bonding hybrid,
$\psi_{\pm}=(\psi_{\rm L}\pm \psi_{\rm R})/\sqrt{2}$ with energies
$E_{\pm}(\vc{k})$ (see Fig.~\ref{fig:Local_SF_states}). The coupling
of these two states manifests in a splitting between the two inner
rings of the Fermi surfaces, decreasing with increasing film thickness
since the overlap of the two surface states decreases exponentially.
In the sections on the lifetime calculation we show results on the
innermost surface-state ring at $k_x>0$, $k_y=0$, noting that the
lifetime $\tau_{\vec k}$ changes only very weakly for different $\vec
k$ corresponding to the same surface-state ring, due to the isotropic
character of the surface states.

In the limit of infinite film thickness only one of the two surface
bands survives at each surface (either $\psi_{\rm L}$ or $\psi_{\rm
  R}$). However, the overlap of these states with the surface-impurity
is higher by a factor $\sqrt{2}$ compared to the overlap of
$\psi_{\pm}$ with the impurity. Therefore, for comparison with
experiments performed not on ultrathin films but on crystal surfaces,
the scattering rates which are calculated here in the $\psi_{\pm}$
basis can be taken from the thick but finite film calculations but
have to be multiplied by a factor 2.

\subsection{Testing of the lifetimes via the optical theorem}
\label{sec:opti_theorem}
Contrary to the case of bulk systems where momentum-relaxation times
and residual resistivities due to scattering at impurities have been
calculated in \emph{ab initio} calculations (e.g.~in
Ref.~\onlinecite{Mertig82,Graefenstein88,Spinflip_PhysRevB_2008}),
there are no numerical or experimental references to compare our
results of lifetime reduction of surface states caused by adatom
scattering.  However, another possibility to attest the numerical
correctness of the results is given by the optical theorem, according
to which the elements of the scattering matrix $T_{\vec k \vec k'}$
must obey
 \begin{equation}
   - \frac{2 N c}{\hbar} \mathbf{\mathrm{Im}} ~T_{\vec k \vec k}   =\frac{2 \pi N^2 c }{V_{\mathrm{BZ}}\hbar}\int_{S(E_{\vec
     k})} \frac{ dS_{\vec k'} }{\hbar v_{\vec k'}} \;\left| T_{\vec k \vec k'}
\right|^2  \equiv \tau_{\vec k}^{-1}
\label{eqn:opt_theo}
\;\mbox{.}
\end{equation}
This identity turns out to be a very sensitive consistency
probe. Also, it is numerically fulfilled only if the numerical
convergence of the quantities involved is very good.  We have
calculated surface-state scattering rates $\tau_{\vec k}^{-1}$ for
impurities in the first surface layer as well as in adatom position on
top of the surface at the fcc threefold-hollow site on a 6 layers
Cu(111) film via both sides of Eq.~(\ref{eqn:opt_theo}).  The
deviation between the left and the right side of
Eq.~(\ref{eqn:opt_theo}) was in all cases less than $1.5\%$ (the
latter deviation was obtained for surface-state scattering rates off
impurities in adatom position in FP calculations). The accuracy of the
calculations is very sensitive to the number of $\vec k$ points used
for the calculation of the structural Green function of the host,
$G_{LL'}^{nn'}(E)$, which is used to calculate the impurity Green
function $G_{LL'}^{{\rm imp}, nn'}(E)$ via a Dyson
equation. Additionally, four vacuum layers are necessary to obtain the
above mentioned accuracy.

\subsection{Comparison of ASA \textit{vs}.~FP and single-site
  \textit{vs}.~multiple-site calculations
\label{sec:comp_ASA_FP}}

The computational effort in ASA is lower than in FP
calculations. However, at surfaces, where the symmetry is broken,
non-spherical components of the potential could play an important
role. Therefore, for the example of a film of 6 layers of copper,
surface-state lifetimes within ASA and FP calculations are calculated
and compared to each other.

\begin{figure*}[!t]
  \begin{center}
       \includegraphics*[width=0.9\textwidth]
         {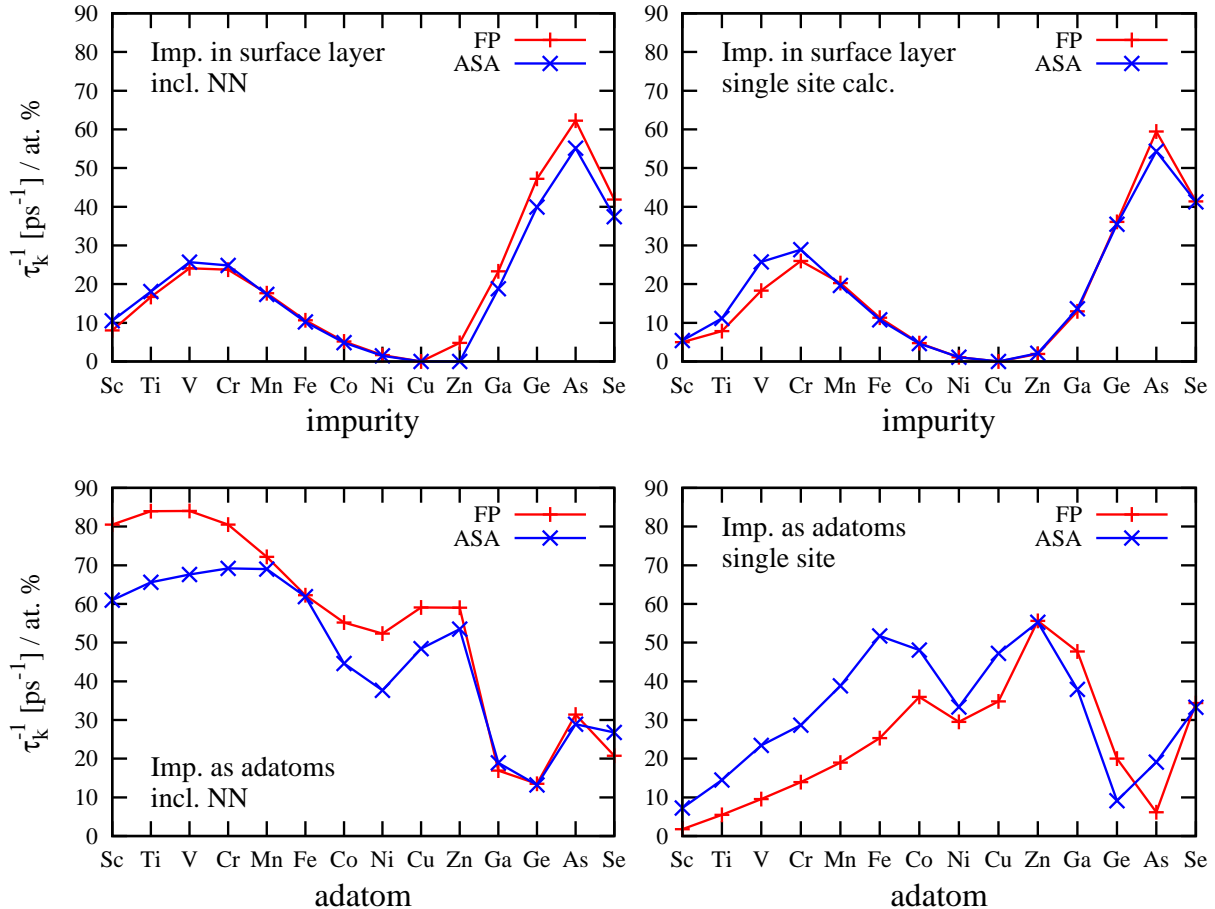}
         \caption{Scattering rates $\tau_{\vec k}^{-1}$ for impurities
           in the first surface layer (\emph{upper panels}) and
           adatoms (\emph{lower panels}) obtained within the atomic
           sphere approximation and full-potential calculations. While
           the agreement between the two calculations is very good for
           impurities in the first surface layer, for adatoms the two
           calculation schemes lead to qualitatively similar but
           quantitatively different results. In the left panels
           results of multiple-site calculations are shown, i.e., with
           a cluster of the nearest neighbors included in the
           potential perturbation. In the right panelsthe comparison
           is shown for single-site calculations.}
  \label{fig:Comp_ASA_FP_slabs_SF}
  \end{center}
\end{figure*}

The results are presented in Fig.~\ref{fig:Comp_ASA_FP_slabs_SF}. In
the two panels on the left the self-consistent perturbed
potential of the impurity and the nearest neighbors (13 sites in
total, which we call a \emph{multiple-site} calculation) has been
included in the calculation of the scattering matrix $T_{\vec k \vec
  k'}$. In the two panels on the right, calculations in the
\emph{single-site-approximation} are presented, i.e.~only
the impurity site is taken in the summation of
Eq.~(\ref{eqn:T_kk_T_nn}), but with the impurity potential previously
converged using also the nearest-neighbors perturbation. All
scattering rates, i.e.~inverse lifetimes $\tau_{\vec k}^{-1}$, are
given in units of ps$^{-1}$ per atomic percent of impurities.
 
For impurities in the first surface layer (the results are shown in
the two upper panels of Fig.~\ref{fig:Comp_ASA_FP_slabs_SF}), the
agreement between the ASA and FP schemes is very good, both for
single-site and multiple-site calculations. This is not the case for
adatoms as can be observed in the two lower panels of the same figure
where a quantitative difference of maximal $25\%$ can be
observed. Nevertheless, the trends in FP and ASA are qualitatively
similar. Since we are interested in these trends, we will restrict our
further calculations to the ASA.

The qualitative difference between single-site multiple-site
calculations in the case of adatoms raises the question whether the
cluster consisting of nearest neighbors is large enough to obtain
converged results. Therefore, multiple-site calculations (not shown
here) including additionally the shells up to the third nearest
neighboring sites (43 atoms in total) have been performed. They show
that the results remain practically stable beyond the nearest-neighbor
approximation, which we adopt henceforth.

\section{Analysis of surface-state lifetimes of C\lowercase{u} films
\label{sec:analy_SF_lifetimes}}

The most striking feature of our results is the qualitative difference
between the trends of scattering rates at adatoms and those of
impurities in the surface, as can be seen in
Fig.~\ref{fig:Comp_ASA_FP_slabs_SF}. While the trend for impurities in
the first surface layer (lower panel) resembles that which has been
found for scattering at impurities in
bulk,\cite{Mertig82,Graefenstein88} the trend for the scattering rate
at adatoms (upper panel) does not.

\begin{figure}[!t]
  \begin{center}
       \includegraphics*[width=0.35\textwidth,angle=-90]
         {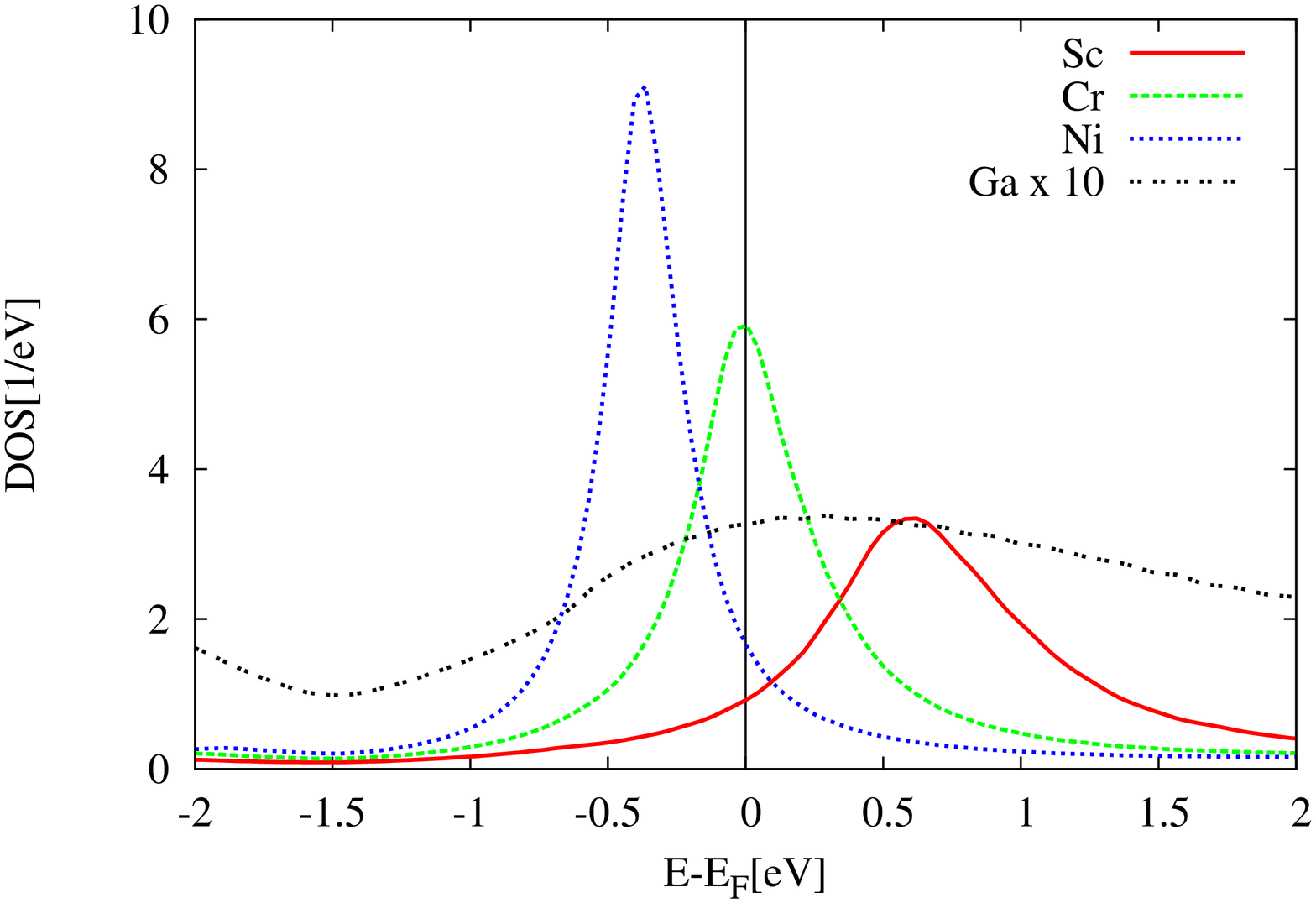}
       \includegraphics*[width=0.35\textwidth, angle=-90]
         {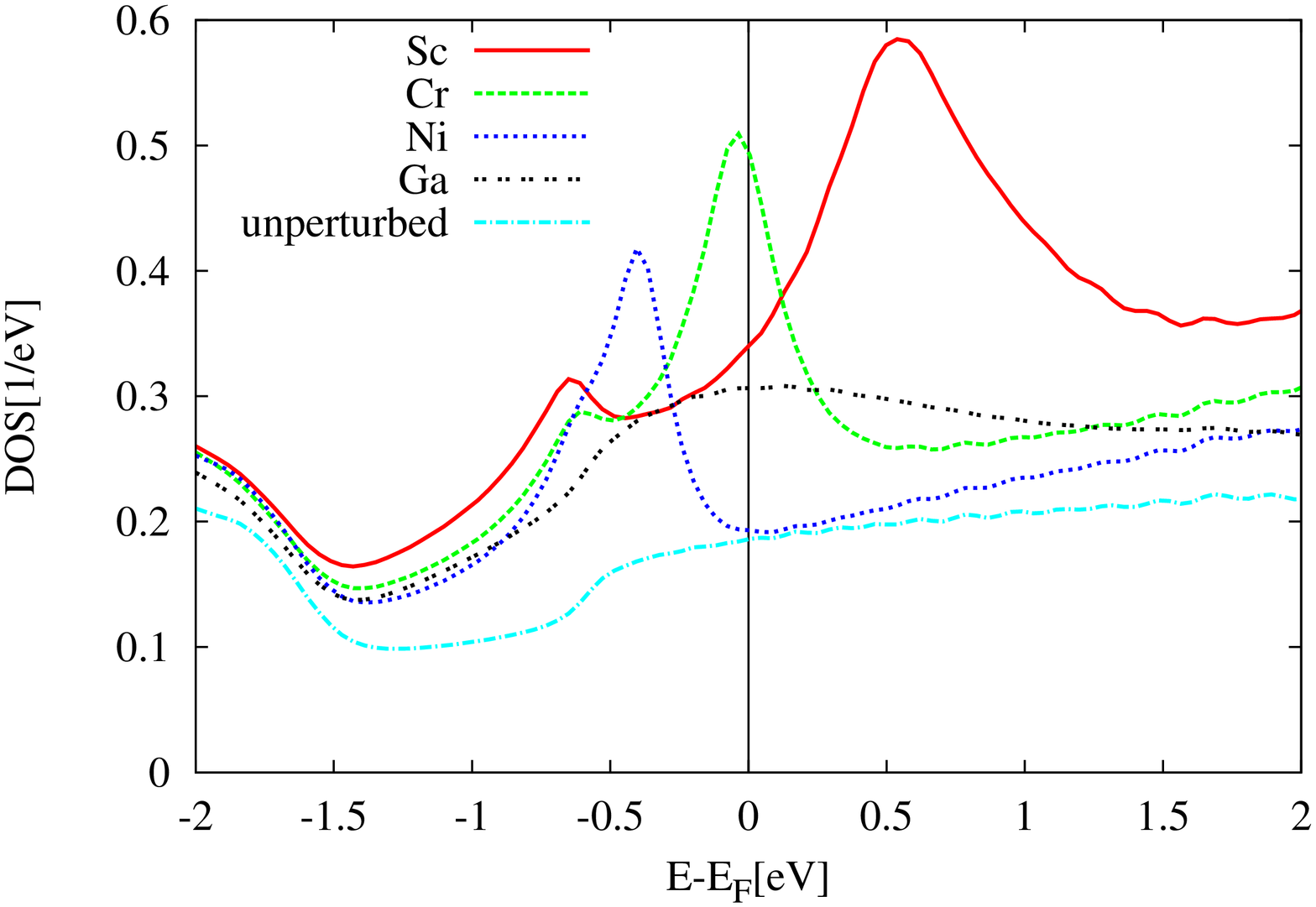}
         \caption{\emph{Top:} Local density of states of Sc, Cr, Ni
           and Ga adatoms on a Cu(111) film of 40 layers. The density
           of states does not differ qualitatively from that of
           impurities in the surface layer. \emph{Bottom:} Density of
           states summed over the vacuum sites surrounding the
           adatoms.}
  \label{fig:Dos_adatoms}
  \end{center}
\end{figure}

In the case of impurities in the first surface layer as well as in
bulk, the observed trend follows the local density of states of the
impurities (see Fig.~\ref{fig:Dos_adatoms}): For the $d$ scatterers,
the scattering rate shows a peak with a maximum at V and Cr as the $d$
resonance crosses the Fermi energy. Then, as the $p$ states start
crossing \EF, the scattering rate increases again starting off
quadratically according to Linde's rule.\cite{Linde_1931,Linde_1932}

\begin{figure}[!t]
  \begin{center}
       \includegraphics*[width=0.35\textwidth, angle=-90]
         {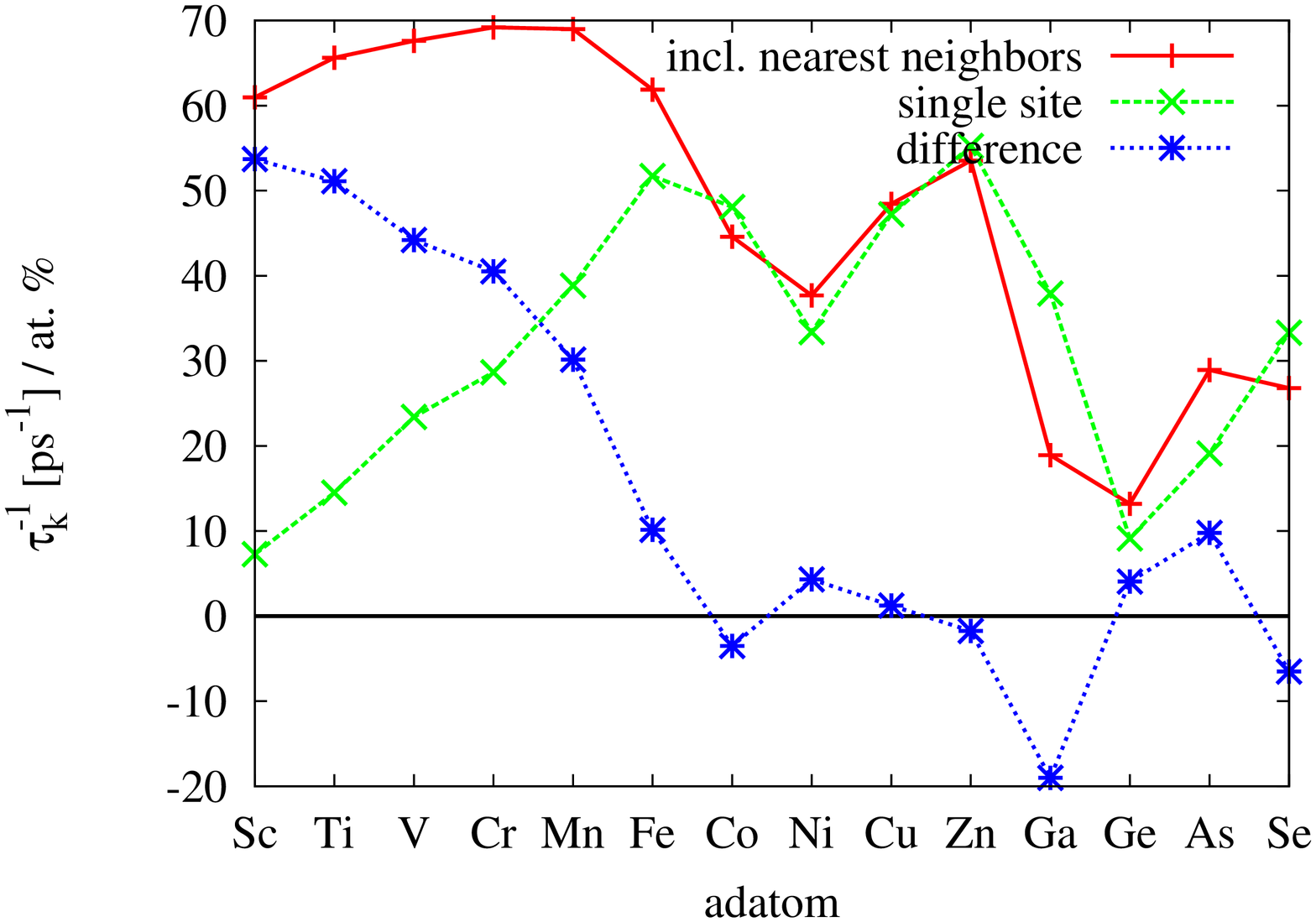}
       \includegraphics*[width=0.35\textwidth, angle=-90]
         {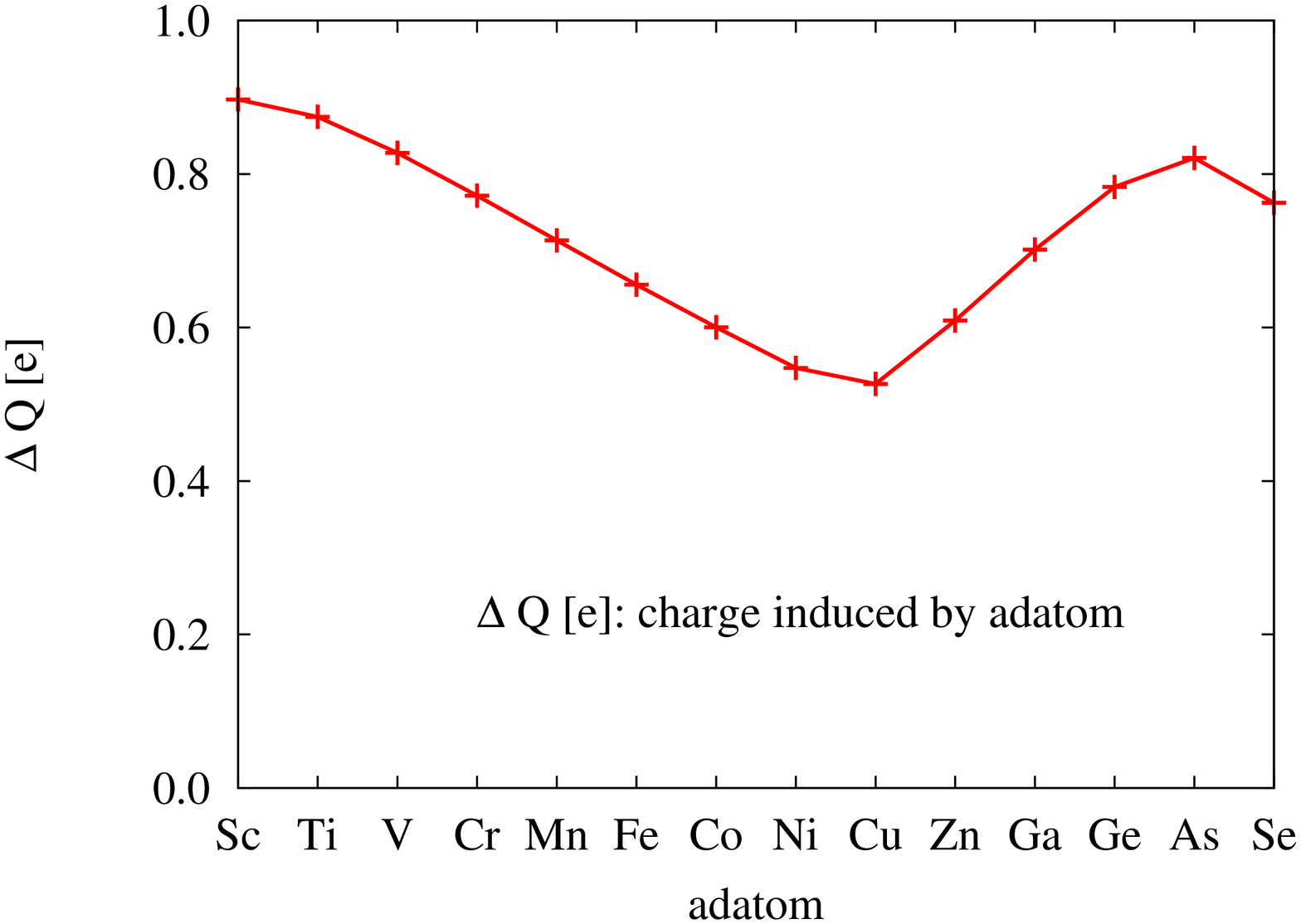}
         \caption{\emph{Top:} Surface-state scattering rate
           $\tau_{\vec k}^{-1}$ in ps$^{-1}$ per atomic percent for
           scattering at adatoms. Red curve: Multiple-site
           calculations. Green curve: Single-site approximation.  Blue
           curve: Difference between the two cases.  \emph{Bottom:}
           Excess charge (in electron-charge units) induced by adatoms
           in the surrounding vacuum sites. For the $d$ adatoms, the
           trend observed in the difference between the scattering
           rates calculated including nearest neighbors and
           single-site scattering is similar to the trend of the
           charge induced by the adatoms.}
  \label{fig:Charge}
  \end{center}
\end{figure}

Basically, this correlation between the impurity local density of
states and the scattering strength of impurities can be traced back to
conditions which are not necessarily met for adatoms. The first
condition is that the screening of the impurity charge in metals takes
place mostly in the impurity atomic cell. Thus the perturbation of the
potential in nearby cells is rather small, while the impurity
wavefunctions are relatively atomic-like in the cell and matched to
phase-shifted host waves outside. The second condition is that the
free-electron-like behavior in the bulk of normal metals together with
the nearly spherical symmetry of the impurity potential allows for
independent scattering of the $s$, $p$ and $d$ without mutual
interference. Thus the angular-momentum resolved local density of
states reflects the scattering properties of the angular-momentum
channels independently to each other.

For adatoms, however, the screening is not as efficient, thus the
atomic size can increase significantly. Here, by atomic size we mean
the volume in which the impurity induces a significant potential
perturbation. In addition, the reduced symmetry allows for
interference between different angular momenta, notably (as we will
see) between $s$ and $p_z$. Impurities in the surface layer are
apparently close enough to the impurity-in-bulk situation that they
behave in the same way.

Having in mind these comments, we proceed to an analysis of the results.

\subsection{Adatom scattering}

Considering the trend of adatom-scattering rates in the multiple-site
approach (see the left-bottom panel of
Fig.~\ref{fig:Comp_ASA_FP_slabs_SF}), there are two features to
observe: The first part of the trend (from Sc to Fe/Co) is dominated
by a relatively large scattering rate, showing a plateau which is
absent in the single-site calculations (right-bottom panel of
Fig.~\ref{fig:Comp_ASA_FP_slabs_SF}). On the other hand, the second
part of the curve (from Ni to Se) is qualitatively the same for
single-site and multiple-site calculations. However, it behaves in an
uncommon manner, showing a kink-like structure, which is completely
different from the trend for the $sp$ impurities in the surface.

The plateau in the scattering rates of the early $3d$ scatterers can
be understood by the significant difference between the single-site
approximation compared to multiple-site calculations (see
Fig.~\ref{fig:Charge}), which hints at an atomic size effect. These
atoms (Sc, Ti, V) have relatively large atomic radii (which are
$r_{\mathrm{at}}=1.62 ~\mathrm{\mathring{A}}$ for Sc, $1.45~
\mathrm{\mathring{A}}$ for Ti, $1.34 ~\mathring{\mathrm{A}}$ for
V)\cite{periodic_table} and therefore extend much more into the vacuum
than e.g.~a Ni adatom with $r_{\mathrm{at}}=1.24
~\mathrm{\mathring{A}}$. This is not any more the case for atoms
embedded in the surface or in bulk, as the strong screening strongly
reduces the atom size.  In order to verify this line of arguments we
have performed additional calculations where -- apart from the adatom
site -- only the perturbed surrounding vacuum sites or only the
perturbed substrate Cu atoms are included in the expression for
$T_{\vec k \vec k'}$ (see the summation over sites $n,n'$ in
Eq.~(\ref{eqn:T_kk_T_nn})). These calculations reveal that the largest
contribution next to the adatom-site arises from the surrounding
vacuum potential, and \emph{not} from the Cu atoms in the surface
layer. We also examined the density of states at \EF\ around the
impurity, shown in the lower panel of Fig.~\ref{fig:Dos_adatoms}. We
found that, indeed, there is an increased DOS in the vacuum around the
early $3d$ adatoms, while this contribution is reduced for the late
$3d$ series. When the $p$ states start crossing \EF\ (at Ga) the size
of the atom grows again, but apparently the qualitative trend is
governed by different effects (see next paragraph), although there are
some quantitative differences between single- and multiple-site
calculations.  The larger extent of the early $3d$ adatoms also
manifests in a larger charge transfer to the surrounding vacuum
sites. This becomes visible in the lower panel of
Fig.~\ref{fig:Charge}, where we show the charge $\Delta Q$ in the
surrounding vacuum sites induced by the adatom.  For the $3d$ adatoms
we find a decreasing trend for $\Delta Q$ similar to the difference
between the scattering rates including the nearest neighbors and
single-site calculations (compare the dotted line of the upper panel
with the lower panel of Fig.~\ref{fig:Charge}). We note that, in the
spirit of the self-consistent density-functional approach, the excess
charge causes extra scattering only through the induced change in the
potential.

\begin{figure}[]
  \begin{center}
       \includegraphics*[width=0.35\textwidth, angle=-90]{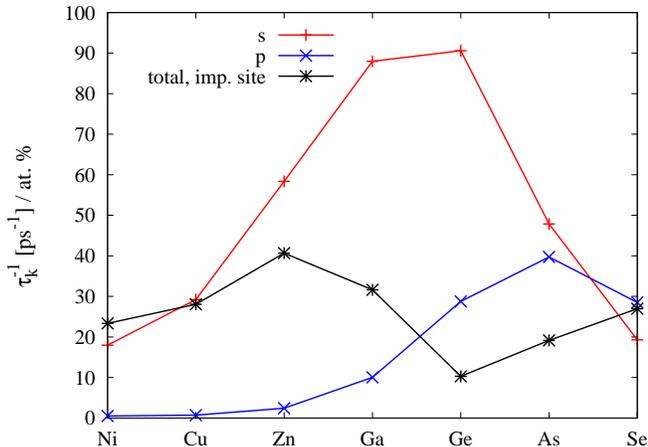}
       \caption{Angular-momentum resolved scattering rates for adatoms
         on a 40 layer Cu(111) film. The inclusion of just the $s$ and
         the $p_z$ channel in the single-site scattering matrix
         $T^{00}_{LL'}$ (compare Eq.~(\ref{eqn:def_T_LLnn})) for the
         $4sp$ adatoms is already a good approximation for the total
         scattering rate. Scattering in only the $s$ channel by far
         overestimates the total scattering rate. The same can be
         observed for Ge, As and Se adatoms when scattering is
         restricted to the $p$ channel. The total scattering rate is
         then reduced due to a destructive interference of $s$ and $p$
         scattering, which can be verified by the investigation of
         Friedel oscillations (see Fig.~\ref{fig:Friedel}).}
  \label{fig:Mom_resolv}
  \end{center}
\end{figure}
\begin{figure}[]
  \begin{center}
       \includegraphics*[width=0.45\textwidth]
         {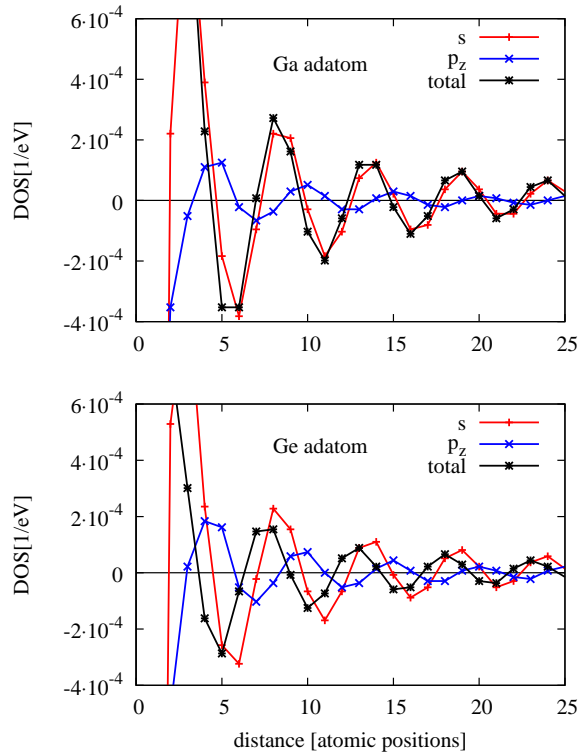}
         \caption{Friedel oscillations in the vacuum near the surface
           and as a function of the distance from the impurity
           calculated for Ga and Ge adatoms.  The difference of the
           local density of states as a function of distance to the
           adatom site is calculated, i.e.~$n^{\rm
             imp}_{L}(E_{\mathrm{F}})-n^{\mathrm{host}}_{L}(E_{\mathrm{F}})$,
           with $n^{\mathrm{host}}_{L}(E_{\mathrm{F}})$ the density of
           states of the vacuum site of the host and $n^{\rm
             imp}_{L}(E_{\mathrm{F}})$ the density of states of the
           same site perturbed by the adatom. Here, the index $L$
           means that the scattering has been artificially confined to
           the $L$ channel only ($s$ or $p_z$). Both for Ga and Ge
           adatoms a destructive interference of the $s$ and $p_z$
           channel is observed as they are out of phase. $d$
           scattering is found to be very small and can be neglected.}
  \label{fig:Friedel}
  \end{center}
\end{figure}

In order to understand the second part of the trend for scattering at
adatoms (from Ni up to Se), which shows a zig-zag structure, we
analyze the single-site results, because qualitatively they do not
differ from multiple-site results, while they allow for an
angular-momentum resolved investigation of scattering rates.  In the
spirit of such an $lm$-decomposition, we restrict the single-site
contribution $n=n'=0$ of the scattering matrix $T_{LL'}^{nn'}$ defined
in Eq.~(\ref{eqn:def_T_LLnn}) to different $L$ channels. The results
are presented in Fig.~\ref{fig:Mom_resolv} for $T_{ss}^{00}$ and
$T_{pp}^{00}$ together with the total single-site contribution. Due to
the presence of the surface, which breaks the symmetry, we expect that
interference is possible among $s$ and $p_z$ waves (the $z$ direction
is taken along the surface normal). The inclusion of the $s$ and the
$p_z$ channel already leads to a curve very close to the total
single-site result ($d_{z^2}$ waves are insignificant for $sp$
adatoms). However, restriction of $T^{00}$ to the $s$ channel only
overestimates the scattering rate for adatoms from Zn to As, while
restriction to the $p$ channel only overestimates the scattering rate
for Ge, As and Se adatoms. The conclusion is that there is destructive
interference between the $s$ and the $p_z$ channels which is allowed
by the breaking of symmetry due to the surface. The zig-zag structure,
with a peak at Zn and a dip at Ge, occurs because the maxima of the
$s$ and $p$ channels are phase-shifted with respect to each other.

For a more elaborate analysis of interference effects we have
calculated Friedel oscillations. In particular we look at the
difference of the local density of states $\Delta n_L(E_{\mathrm{F}})$
in the vaccum close the impurity, and in a direction parallel to the
surface, between the system with impurity and the host system. To this
end we integrated, for the two cases, the density at \EF\ over the
atomic spheres of the vacuum sites with and without the impurity,
obtaining $\Delta n_L(E_{\mathrm{F}})=n^{\rm
  host}_{L}(E_{\mathrm{F}})-n^{\rm imp}_{L}(E_{\mathrm{F}})$, for Ga
and Ge adatoms. The index $L$ here means that we made again an
angular-momentum decomposition, by allowing scattering by only a
certain $L$ channel while setting the others to zero. In this way we
are able to see, in real space, the relative phase shift in channels
which are expected to interfere, e.g., between the $s$ and $p_z$
channel.  This analysis is presented in Fig.~\ref{fig:Friedel} as
function of the distance from the adatom site. We see that
oscillations in the $s$ and $p_z$ channel are mutually phase-shifted
and interfere destructively.

\subsection{Separation of scattering into bulk and surface states}

After a scattering event, an electron that was previously occupying a
surface state $\vc{k}_{\rm surf}$ can end up in another surface state
$\vc{k}'_{\rm surf}$ or in a bulk state $\vc{k}'_{\rm bulk}$. The
distinction between the two cases can be of importance to effects such
as surface-state mediated interactions between defects, surface
resistivity, the lifetime of surface plasmons, etc..  The method used
to calculate the surface-state lifetimes given in section
\ref{sec:theory}, Eq.~(\ref{eqn:inv_lifetime_int}), allows to
distinguish between the two cases. The total scattering rate of a
(surface) state characterized by a wavevector $\vec k$ is composed of
a contribution $1/\tau_{\vec k}^{\mathrm{surf}}$ given by
\begin{equation}
    \frac{1}{\tau_{\vec k}^{\mathrm{surf}}}  
 =  \frac{2 \pi N^2c}{V_{\mathrm{BZ}} \hbar^2}\int_{S(E_\mathrm{F})_{
 \mathrm{surf}}} \frac{dS_{\vec k'}}{v_{\vec k'}}  \left| T_{\vec k \vec k'} \right|^2  
 \;\mbox{,}
\label{eqn:inv_lifetime_SF_S}
\end{equation}
where the integration is performed only over the surface bands and
analogously a contribution $1/\tau_{\vec k}^{\mathrm{bulk}}$, where
only the bulk end-states of the Fermi surface are taken into account:
\begin{equation}
    \frac{1}{\tau_{\vec k}^{\mathrm{bulk}}}  
 =  \frac{2 \pi N^2c}{V_{\mathrm{BZ}} \hbar^2}\int_{S(E_\mathrm{F})_{\mathrm{bulk}}} \frac{dS_{\vec k'}}{v_{\vec k'}}  \left| T_{\vec k \vec k'} \right|^2    \mbox{.}
\label{eqn:inv_lifetime_bulk_S}
\end{equation}
Obviously, considering the definitions of $\tau_{\vec k}$, $\tau_{\vec
k}^{\mathrm{surf}}$ and $\tau_{\vec k}^{\mathrm{bulk}}$, the relation
\begin{equation}
    \frac{1}{\tau_{\vec k}} = \frac{1}{\tau_{\vec k}^{\mathrm{surf}}}+ \frac{1}{
    \tau_{\vec k}^{\mathrm{bulk}}}    
    \label{eqn:sum_SF_BULK}
\end{equation}
holds.

\begin{figure*}[]
  \begin{center}
       \includegraphics*[width=0.9\textwidth]
         {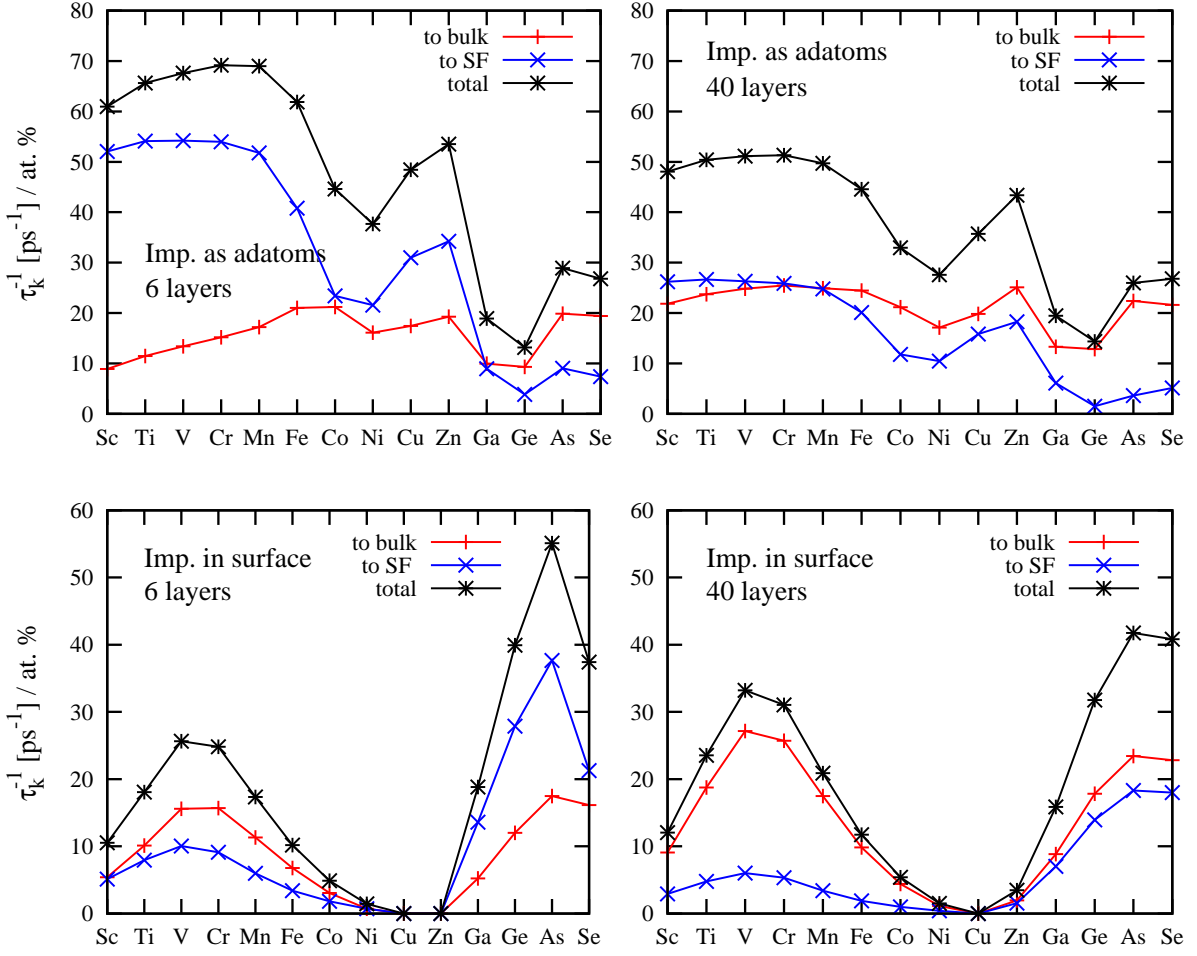}
         \caption{Scattering rate $\tau_{\vec k}^{-1}$ for the
           innermost surface state together with the contributions of
           scattering rates to bulk and surface states $1/\tau_{\vec
             k}^{\mathrm{bulk}}$ and $1/\tau_{\vec k}^{\mathrm{surf}}$
           for films with 6 layers (\emph{left panels}) and 40 layers
           of copper (\emph{right panels}) for adatoms (\emph{top})
           and impurities in the first surface layer
           (\emph{bottom}). }
  \label{fig:Scat_bulk_SF}
  \end{center}
\end{figure*}

The two contributions have been calculated for films consisting of 6
and 40 layers of copper for impurities in the surface layer as well as
for adatoms and are presented in Fig.~\ref{fig:Scat_bulk_SF}. The most
important observation is that $1/\tau_{\vec k}^{\mathrm{bulk}}$ and
$1/\tau_{\vec k}^{\mathrm{surf}}$ are on the same order of magnitude,
even in the case of adatoms. Naturally, variations occur as a function
of film thickness, which we will now address.

Qualitatively, we expect that the relative scattering rate into bulk
and surface states should depend on (i) the amplitude of the surface
states, $\psi_S(\vec{R}_{\rm imp})$, at the impurity site, (ii) the
amplitude of the bulk states, $\psi_B(\vec{R}_{\rm imp})$, at the
impurity site, and (iii) the available number of bulk states,
$N_{\rm B}$. About point (i), as a function of the number of layers in the
film, $N_{\rm L}$, $\psi_S(\vec{R}_{\rm imp})$ first decreases before
reaching saturation, since in very thin films the number of layers in
which the surface states can penetrate is limited (see also
Fig.~\ref{fig:Local_SF_states}). From this effect alone we expect an
overall reduction of both $1/\tau_{\vec k}^{\mathrm{surf}}\sim
|\langle\psi_S(\vec{R}_{\rm imp})|T|\psi_S(\vec{R}_{\rm
  imp})\rangle|^2\sim |\psi_S(\vec{R}_{\rm imp})|^4$ and $1/\tau_{\vec
  k}^{\mathrm{bulk}}\sim |\langle\psi_S(\vec{R}_{\rm
  imp})|T|\psi_B(\vec{R}_{\rm imp})\rangle|^2\sim |\psi_S(\vec{R}_{\rm
  imp})|^2$ with increasing $N_{\rm L}$ before a saturation takes
place, but with $1/\tau_{\vec k}^{\mathrm{surf}}$ being more strongly
affected due to the 4-th power of the amplitude. About point (ii),
$\psi_B(\vec{R}_{\rm imp})\sim 1/\sqrt{N_{\rm L}}$, yielding a factor
of $1/N_{\rm L}$ to $1/\tau_{\vec k}^{\mathrm{bulk}}$; however, this
reduction is partly compensated by point (iii), which contributes a
factor $N_{\rm B}=N_{\rm L}-1$ to $1/\tau_{\vec k}^{\mathrm{bulk}}$.

Combining (i), (ii) and (iii) gives a behavior of the form
$1/\tau_{\vec k}^{\mathrm{bulk}}\sim |\psi_S(\vec{R}_{\rm
  imp})|^2(1-1/N_{\rm L})$ for scattering into bulk states, which
should show a saturation for large $N_{\rm L}$, but could be either
increasing or decreasing at small $N_{\rm L}$ depending on the
relative importance of the two terms. On the other hand, point (i)
gives a behavior of the form $1/\tau_{\vec k}^{\mathrm{surf}}\sim
|\psi_S(\vec{R}_{\rm imp})|^4$ for scattering into surface states,
which should be first decreasing with $N_{\rm L}$ and then also
saturating.

The qualitatively expected behavior is consistent with our
calculations (Fig.~\ref{fig:Scat_bulk_SF}).  The scattering rates into
surface states are larger for the 6 layer film than for the 40 layer
film by approximately a factor 2, while the scattering rates into
bulk states are a little larger for the 40 layer film.

\section{Comparison between C\lowercase{u}, A\lowercase{g} and A\lowercase{u} films
  \label{sec:SF_tau_Cu_Ag_Au}}

In the previous section we have analyzed surface-state lifetimes for
impurities in and on copper films. Here we extend our investigations
to silver and gold films. Even though the band structure of the three
noble metals is similar, differences are anticipated mainly due to the
change of energetic position of the surface state, especially in Ag.

\begin{figure*}[]
  \begin{center}
       \includegraphics*[width=0.8\textwidth]
         {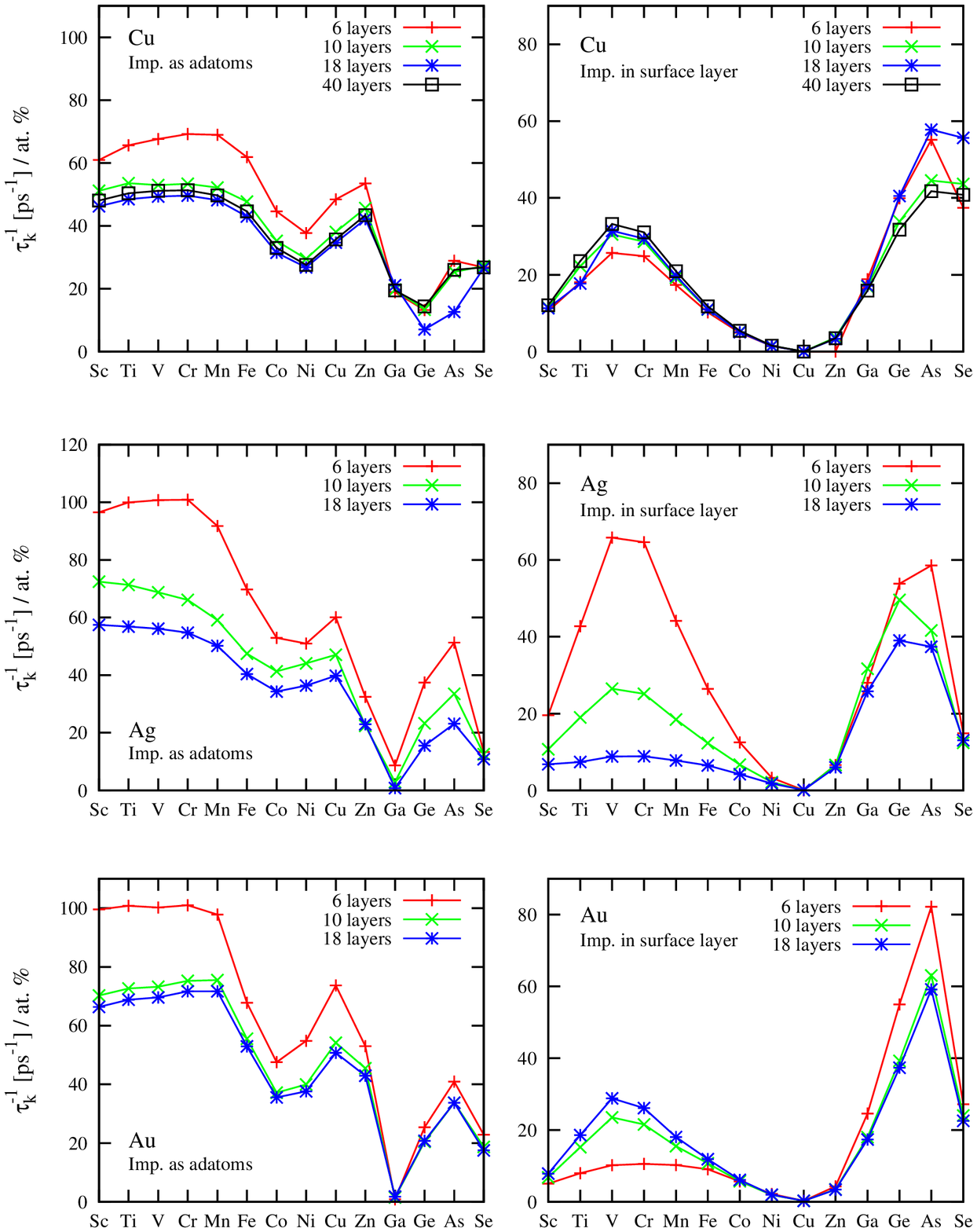}
         \caption{Surface-state scattering rate $\tau_{\vec k}^{-1}$
           for adatoms (\emph{left panels}) and impurities in the
           first surface layer (\emph{right panels}) for Cu, Ag and Au
           films with different number of layers. For silver films the
           scattering rate is most sensitive to the film
           thickness due to the fact that one of the surface bands
           lies above \EF\ in the 6 and 10 layer cases. Qualitatively
           the trend as a function of the impurity atomic number is
           similar for the three hosts.}
  \label{fig:Scat_CuAgAu}
  \end{center}
\end{figure*}

The calculated data, shown in Fig.~\ref{fig:Scat_CuAgAu}, are
qualitatively similar for all three hosts. Scattering rates off
adatoms are largest for the first elements of the row and remain
almost constant until Mn; just as in the case of the copper film (see
previous section), this should be due to the larger atomic radius
entailing a larger extent of these adatoms into the
vacuum. Furthermore, all three host materials show a clear trend that
scattering at adatoms on a film of six layers is enhanced compared to
that of larger thickness. The reason is the higher
localization of the surface state for thin films, as explained in the
Cu(111) case. 

We show here results for the innermost surface state only, whose
lifetime, especially for the 6 layer films and for some adatoms,
remarkably differs from that of the second surface state. While for
the early $3d$ elements the difference is relatively small (a few
percent), it can increase up to $60 \%$ for Ge and $48 \%$ for As
adatoms. For the 40 layer film, this difference almost vanishes as
does the splitting between the two surface bands.

Concerning scattering at impurities in the surface, the situation is
somewhat complicated. Qualitatively, for all three host materials a
clear maximal scattering rate for the $3d$ elements with half-filled
shells (V, Cr) is observed as well as high scattering rates for the
$sp$ scatterers.  Hence, as expected, the global trend reflects the
situation of scattering at impurities in the bulk. However, large
quantitative differences among the three host materials are observed
when considering the thickness-dependence as well as the comparative
scattering strength of $3d$ and $4sp$ impurities. While for a copper
host, the scattering rate does not depend much on the film thickness,
especially for silver films the thickness can change the scattering
rate up to almost an order of magnitude.  Actually, such a strong
variation in the Ag films is reasonable, because of the relatively
shallow position of the surface state with respect to the Fermi
energy.\cite{footnote2} As a result, in thin films quantum-confinement
effects push one of the two surface states (the innermost ring) above
$E_{\mathrm{F}}$ and the other lower, resulting in a significant
change of the available phase space for scattering. In thicker films
the two states are approximately degenerate and below \EF. In any
case, in Ag the scattering rate into surface states is small due to
the smallness of the corresponding Fermi rings.

A comparison to other theoretical or experimental results is not
possible because of lack of data; although these surface states have
been subject of many experiments, to our knowledge no experiments have
been performed in which the surface-state lifetimes due to scattering
at the specific impurities at the Fermi level have been
measured. However, the order of magnitude of the calculated scattering
rates should allow for an experimental detection, which e.g. in ARPES
reaches a few meV. To compare, in inverse photoemission spectroscopy a
linewidth of about $23~ \mathrm{meV} \approx 35 ~\mathrm{ps}^{-1}
\hbar$ for Cu, $6 ~\mathrm{meV} \approx 9 ~\mathrm{ps}^{-1} \hbar$ for
Ag and $21~ \mathrm{meV} \approx 33 ~\mathrm{ps}^{-1} \hbar$ for Au
for electron-electron scattering processes has been
measured.\cite{PhysRevB.63.115415} A measurement for $1\%$ of defects
should be, therefore, within the experimental resolution.

\section{Scattering at magnetic impurities}
So far, only scattering at non-magnetic impurities and adatoms has
been considered, in an approximation that was explained in the
introduction. However, above the Kondo temperature, some of the $3d$
impurities become magnetic and scattering in the two spin channels has
to be treated separately. We investigate now the consequences of spin
polarization, showing results for the Cu(111) surface only, which
suffices to demonstrate the main effect of a double-peak structure in
the scattering rate.

\begin{figure*}[!t]
  \begin{center}
       \includegraphics*[width=0.9\textwidth]
         {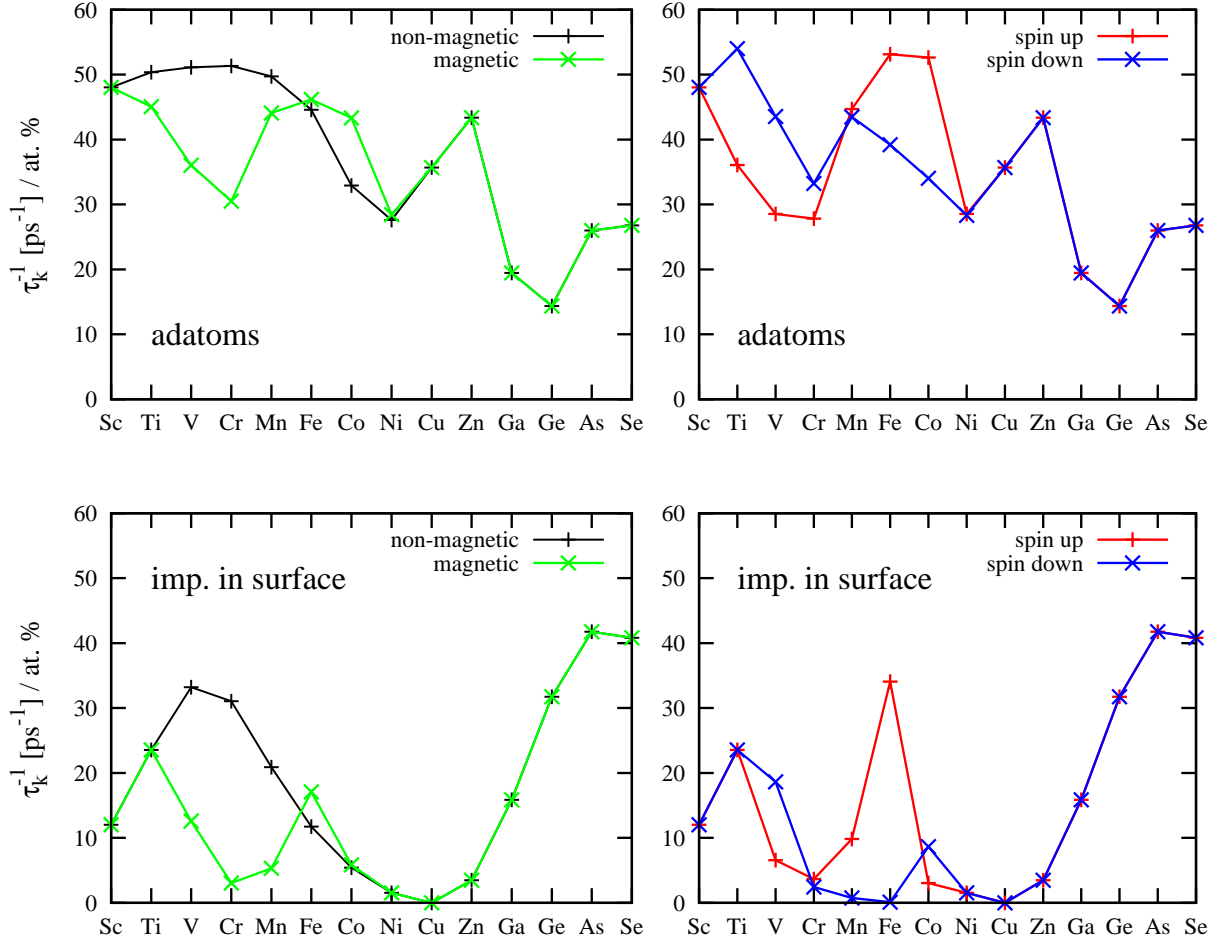}
         \caption{Scattering rate of the surface state at magnetic
           adatoms (\emph{top}) and impurities in the surface layer
           (\emph{bottom}) of a 40 layer Cu(111) film. The spin polarization
           of the $3d$ elements leads to a double-peak structure of
           $\tau_{\vec k}^{-1}$ because of the exchange splitting in the density
           of states of the two spin channels. The first maximum is
           reached for Ti, when the $d$ resonance of the spin-up
           channel crosses the Fermi level, while the second maximum
           corresponds to the localization of the $d$ resonance of the
           spin down channel at \EF. In the right panels the
           spin-resolved results are shown.}
  \label{fig:Scat_CuAgAu_magn}
  \end{center}
\end{figure*}

The calculated lifetimes are shown in Fig.~\ref{fig:Scat_CuAgAu_magn}
for a 40-layer film.  In the adatom position the $3d$ elements
starting between Ti and Ni are magnetic, while as impurities in the
surface Ti and Ni are paramagnetic. The $4sp$ elements are in all
cases non-magnetic. The magnetism of the $3d$ elements leads to an
expected double-peak structure in the trend of the scattering rates,
which is already known e.g.~for residual resistivity in bulk above the
Kondo temperature\cite{PhysRevB.22.5777,Gruner74,Cohen78,Coleridge85} and
originates from the offset of the $d$ resonances of the two spin
channels, which are mutually repelled by the exchange
interaction. Thus, a first peak of the scattering rate is observed for
Ti, where the $d$ resonance of the majority channel is centered at the
Fermi level, while a second peak appears for Fe and Co in the
minority-spin channel. For Cr impurities, where scattering rates are
large in the case of paramagnetic impurities, the scattering rate is
most drastically reduced as \EF\ lies between the resonances.

\section{Summary}

In summary, we have presented a systematic study of the lifetime of
the (111) surface state of the noble metals considering scattering at
the Fermi energy due to $3d$ and $4sp$ impurities atoms on and in the
surface.

Our main finding is a qualitatively different trend, as a function of
the impurity atomic number, for adatoms compared to surface-embedded
impurities. While the latter case is similar to the trend for
impurities in the bulk, reflecting the sharpness of the $d$-resonance
and Linde's rule as the $d$ and $p$ states cross the Fermi level,
adatom scattering is found to be dominated by other criteria such as
the size of the adatom and interference effects between the scattering
amplitudes of different angular momenta. The conclusion is that the
surface strongly affects the scattering properties by the reduction in
charge screening as well as by the lowering of symmetry.

A separation of scattering rates into surface-to-surface and
surface-to-bulk parts shows that the two contributions are comparable
in magnitude for adatoms as well as for embedded
impurities. Additionally, in the case of ultrathin films we have found
that the scattering into surface-states contribution decreases with
film thickness, since the surface state extends further into the bulk
in thicker films, and finally saturates. Furthermore, we have found
that the lifetime is sensitive on the film thickness, especially for
Ag. This is due to the very shallow surface state of Ag, which can be
shifted in energy due to quantum confinement effects, changing the
available phase space for scattering.

The calculated lifetimes should be detectable as a surface-state
linewidth within the resolution of photoemission spectroscopy already
for impurity concentrations 1\%.

\section*{Acknowledgments}
We are indebted to Prof. P. H. Dederichs for valuable discussions.
S.~L. acknowledges the support of the HGF-YIG Programme VH-NG-717.


\begin{thebibliography}{45}
\expandafter\ifx\csname natexlab\endcsname\relax\def\natexlab#1{#1}\fi
\expandafter\ifx\csname bibnamefont\endcsname\relax
  \def\bibnamefont#1{#1}\fi
\expandafter\ifx\csname bibfnamefont\endcsname\relax
  \def\bibfnamefont#1{#1}\fi
\expandafter\ifx\csname citenamefont\endcsname\relax
  \def\citenamefont#1{#1}\fi
\expandafter\ifx\csname url\endcsname\relax
  \def\url#1{\texttt{#1}}\fi
\expandafter\ifx\csname urlprefix\endcsname\relax\def\urlprefix{URL }\fi
\providecommand{\bibinfo}[2]{#2}
\providecommand{\eprint}[2][]{\url{#2}}

\bibitem[{\citenamefont{Kr{\"o}ger et~al.}(2007)\citenamefont{Kr{\"o}ger,
  Becker, Jensen, von Hofe, N{\'e}el, Limot, Berndt, Crampin, Pehlke, Corriol
  et~al.}}]{Kroeger_STM_surface_states}
\bibinfo{author}{\bibfnamefont{J.}~\bibnamefont{Kr{\"o}ger}},
  \bibinfo{author}{\bibfnamefont{M.}~\bibnamefont{Becker}},
  \bibinfo{author}{\bibfnamefont{H.}~\bibnamefont{Jensen}},
  \bibinfo{author}{\bibfnamefont{T.}~\bibnamefont{von Hofe}},
  \bibinfo{author}{\bibfnamefont{N.}~\bibnamefont{N{\'e}el}},
  \bibinfo{author}{\bibfnamefont{L.}~\bibnamefont{Limot}},
  \bibinfo{author}{\bibfnamefont{R.}~\bibnamefont{Berndt}},
  \bibinfo{author}{\bibfnamefont{S.}~\bibnamefont{Crampin}},
  \bibinfo{author}{\bibfnamefont{E.}~\bibnamefont{Pehlke}},
  \bibinfo{author}{\bibfnamefont{C.}~\bibnamefont{Corriol}},
  \bibnamefont{et~al.}, \bibinfo{journal}{Progress in Surface Science}
  \textbf{\bibinfo{volume}{82}}, \bibinfo{pages}{293} (\bibinfo{year}{2007}).

\bibitem[{\citenamefont{Manoharan et~al.}(2000)\citenamefont{Manoharan, Lutz,
  and Eigler}}]{Manoharan00}
\bibinfo{author}{\bibfnamefont{H.~C.} \bibnamefont{Manoharan}},
  \bibinfo{author}{\bibfnamefont{C.~P.} \bibnamefont{Lutz}}, \bibnamefont{and}
  \bibinfo{author}{\bibfnamefont{D.~M.} \bibnamefont{Eigler}},
  \bibinfo{journal}{Nature (London)} \textbf{\bibinfo{volume}{403}},
  \bibinfo{pages}{512} (\bibinfo{year}{2000}).

\bibitem[{\citenamefont{Vitali et~al.}(2003)\citenamefont{Vitali, Wahl,
  Schneider, Kern, Silkin, Chulkov, and Echenique}}]{Vitali2003L47}
\bibinfo{author}{\bibfnamefont{L.}~\bibnamefont{Vitali}},
  \bibinfo{author}{\bibfnamefont{P.}~\bibnamefont{Wahl}},
  \bibinfo{author}{\bibfnamefont{M.~A.} \bibnamefont{Schneider}},
  \bibinfo{author}{\bibfnamefont{K.}~\bibnamefont{Kern}},
  \bibinfo{author}{\bibfnamefont{V.~M.} \bibnamefont{Silkin}},
  \bibinfo{author}{\bibfnamefont{E.~V.} \bibnamefont{Chulkov}},
  \bibnamefont{and} \bibinfo{author}{\bibfnamefont{P.~M.}
  \bibnamefont{Echenique}}, \bibinfo{journal}{Surface Science}
  \textbf{\bibinfo{volume}{523}}, \bibinfo{pages}{L47 } (\bibinfo{year}{2003}).

\bibitem[{\citenamefont{McDougall et~al.}(1995)\citenamefont{McDougall,
  Balasubramanian, and Jensen}}]{PhysRevB.51.13891}
\bibinfo{author}{\bibfnamefont{B.~A.} \bibnamefont{McDougall}},
  \bibinfo{author}{\bibfnamefont{T.}~\bibnamefont{Balasubramanian}},
  \bibnamefont{and} \bibinfo{author}{\bibfnamefont{E.}~\bibnamefont{Jensen}},
  \bibinfo{journal}{Phys. Rev. B} \textbf{\bibinfo{volume}{51}},
  \bibinfo{pages}{13891} (\bibinfo{year}{1995}).

\bibitem[{\citenamefont{Reinert et~al.}(2001)\citenamefont{Reinert, Nicolay,
  Schmidt, Ehm, and H\"ufner}}]{PhysRevB.63.115415}
\bibinfo{author}{\bibfnamefont{F.}~\bibnamefont{Reinert}},
  \bibinfo{author}{\bibfnamefont{G.}~\bibnamefont{Nicolay}},
  \bibinfo{author}{\bibfnamefont{S.}~\bibnamefont{Schmidt}},
  \bibinfo{author}{\bibfnamefont{D.}~\bibnamefont{Ehm}}, \bibnamefont{and}
  \bibinfo{author}{\bibfnamefont{S.}~\bibnamefont{H\"ufner}},
  \bibinfo{journal}{Phys. Rev. B} \textbf{\bibinfo{volume}{63}},
  \bibinfo{pages}{115415} (\bibinfo{year}{2001}).

\bibitem[{\citenamefont{Reinert et~al.}(2004)\citenamefont{Reinert, Eltner,
  Nicolay, Forster, Schmidt, and H√ºfner}}]{Reinert2004229}
\bibinfo{author}{\bibfnamefont{F.}~\bibnamefont{Reinert}},
  \bibinfo{author}{\bibfnamefont{B.}~\bibnamefont{Eltner}},
  \bibinfo{author}{\bibfnamefont{G.}~\bibnamefont{Nicolay}},
  \bibinfo{author}{\bibfnamefont{F.}~\bibnamefont{Forster}},
  \bibinfo{author}{\bibfnamefont{S.}~\bibnamefont{Schmidt}}, \bibnamefont{and}
  \bibinfo{author}{\bibfnamefont{S.}~\bibnamefont{H√ºfner}},
  \bibinfo{journal}{Physica B: Condensed Matter}
  \textbf{\bibinfo{volume}{351}}, \bibinfo{pages}{229 } (\bibinfo{year}{2004}),
  \bibinfo{note}{proceedings of The International Symposium on Synchrotron
  Radiation Research for Spin and Electronic States in d and f Electron
  Systems}.

\bibitem[{\citenamefont{Kr{\"o}ger et~al.}(2005)\citenamefont{Kr{\"o}ger,
  Limot, Jensen, Berndt, Crampin, and Pehlke}}]{progr_surfa_science_Kroeger}
\bibinfo{author}{\bibfnamefont{J.}~\bibnamefont{Kr{\"o}ger}},
  \bibinfo{author}{\bibfnamefont{L.}~\bibnamefont{Limot}},
  \bibinfo{author}{\bibfnamefont{H.}~\bibnamefont{Jensen}},
  \bibinfo{author}{\bibfnamefont{R.}~\bibnamefont{Berndt}},
  \bibinfo{author}{\bibfnamefont{S.}~\bibnamefont{Crampin}}, \bibnamefont{and}
  \bibinfo{author}{\bibfnamefont{E.}~\bibnamefont{Pehlke}},
  \bibinfo{journal}{Progress in Surface Science} \textbf{\bibinfo{volume}{80}},
  \bibinfo{pages}{26} (\bibinfo{year}{2005}).

\bibitem[{\citenamefont{Kevan}(1983)}]{PhysRevLett.50.526}
\bibinfo{author}{\bibfnamefont{S.~D.} \bibnamefont{Kevan}},
  \bibinfo{journal}{Phys. Rev. Lett.} \textbf{\bibinfo{volume}{50}},
  \bibinfo{pages}{526} (\bibinfo{year}{1983}).

\bibitem[{\citenamefont{Limot et~al.}(2005)\citenamefont{Limot, Pehlke,
  Kr\"oger, and Berndt}}]{PhysRevLett.94.036805}
\bibinfo{author}{\bibfnamefont{L.}~\bibnamefont{Limot}},
  \bibinfo{author}{\bibfnamefont{E.}~\bibnamefont{Pehlke}},
  \bibinfo{author}{\bibfnamefont{J.}~\bibnamefont{Kr\"oger}}, \bibnamefont{and}
  \bibinfo{author}{\bibfnamefont{R.}~\bibnamefont{Berndt}},
  \bibinfo{journal}{Phys. Rev. Lett.} \textbf{\bibinfo{volume}{94}},
  \bibinfo{pages}{036805} (\bibinfo{year}{2005}).

\bibitem[{\citenamefont{Li et~al.}(1998)\citenamefont{Li, Schneider, Berndt,
  Bryant, and Crampin}}]{PhysRevLett.81.4464}
\bibinfo{author}{\bibfnamefont{J.}~\bibnamefont{Li}},
  \bibinfo{author}{\bibfnamefont{W.-D.} \bibnamefont{Schneider}},
  \bibinfo{author}{\bibfnamefont{R.}~\bibnamefont{Berndt}},
  \bibinfo{author}{\bibfnamefont{O.~R.} \bibnamefont{Bryant}},
  \bibnamefont{and} \bibinfo{author}{\bibfnamefont{S.}~\bibnamefont{Crampin}},
  \bibinfo{journal}{Phys. Rev. Lett.} \textbf{\bibinfo{volume}{81}},
  \bibinfo{pages}{4464} (\bibinfo{year}{1998}).

\bibitem[{\citenamefont{Kliewer et~al.}(2001)\citenamefont{Kliewer, Berndt, and
  Crampin}}]{1367-2630-3-1-322}
\bibinfo{author}{\bibfnamefont{J.}~\bibnamefont{Kliewer}},
  \bibinfo{author}{\bibfnamefont{R.}~\bibnamefont{Berndt}}, \bibnamefont{and}
  \bibinfo{author}{\bibfnamefont{S.}~\bibnamefont{Crampin}},
  \bibinfo{journal}{New Journal of Physics} \textbf{\bibinfo{volume}{3}},
  \bibinfo{pages}{22} (\bibinfo{year}{2001}).

\bibitem[{\citenamefont{Jensen et~al.}(2005)\citenamefont{Jensen, Kr\"oger,
  Berndt, and Crampin}}]{PhysRevB.71.155417}
\bibinfo{author}{\bibfnamefont{H.}~\bibnamefont{Jensen}},
  \bibinfo{author}{\bibfnamefont{J.}~\bibnamefont{Kr\"oger}},
  \bibinfo{author}{\bibfnamefont{R.}~\bibnamefont{Berndt}}, \bibnamefont{and}
  \bibinfo{author}{\bibfnamefont{S.}~\bibnamefont{Crampin}},
  \bibinfo{journal}{Phys. Rev. B} \textbf{\bibinfo{volume}{71}},
  \bibinfo{pages}{155417} (\bibinfo{year}{2005}).

\bibitem[{\citenamefont{Li et~al.}(1999)\citenamefont{Li, Schneider, Crampin,
  and Berndt}}]{Li199995}
\bibinfo{author}{\bibfnamefont{J.}~\bibnamefont{Li}},
  \bibinfo{author}{\bibfnamefont{W.~D.} \bibnamefont{Schneider}},
  \bibinfo{author}{\bibfnamefont{S.}~\bibnamefont{Crampin}}, \bibnamefont{and}
  \bibinfo{author}{\bibfnamefont{R.}~\bibnamefont{Berndt}},
  \bibinfo{journal}{Surface Science} \textbf{\bibinfo{volume}{422}},
  \bibinfo{pages}{95 } (\bibinfo{year}{1999}).

\bibitem[{\citenamefont{Kliewer et~al.}(2000)\citenamefont{Kliewer, Berndt,
  Chulkov, Silkin, Echenique, and Crampin}}]{Kliewer26052000}
\bibinfo{author}{\bibfnamefont{J.}~\bibnamefont{Kliewer}},
  \bibinfo{author}{\bibfnamefont{R.}~\bibnamefont{Berndt}},
  \bibinfo{author}{\bibfnamefont{E.~V.} \bibnamefont{Chulkov}},
  \bibinfo{author}{\bibfnamefont{V.~M.} \bibnamefont{Silkin}},
  \bibinfo{author}{\bibfnamefont{P.~M.} \bibnamefont{Echenique}},
  \bibnamefont{and} \bibinfo{author}{\bibfnamefont{S.}~\bibnamefont{Crampin}},
  \bibinfo{journal}{Science} \textbf{\bibinfo{volume}{288}},
  \bibinfo{pages}{1399} (\bibinfo{year}{2000}).

\bibitem{footnote1}
The temperature dependence of the contribution of the
  electron-phonon coupling to the surface-state linewidth can be
  addressed by ARPES only, since at elevated temperature STS
  measurements do not lead to significant results; this is due to a
  change of the width of the Fermi distribution function
  \cite{Matzdorf2000151} and thus an intrinsic broadening.

\bibitem[{\citenamefont{Kevan}(1986)}]{PhysRevB.33.4364}
\bibinfo{author}{\bibfnamefont{S.~D.} \bibnamefont{Kevan}},
  \bibinfo{journal}{Phys. Rev. B} \textbf{\bibinfo{volume}{33}},
  \bibinfo{pages}{4364} (\bibinfo{year}{1986}).

\bibitem[{\citenamefont{Theilmann et~al.}(1997)\citenamefont{Theilmann,
  Matzdorf, Meister, and Goldmann}}]{PhysRevB.56.3632}
\bibinfo{author}{\bibfnamefont{F.}~\bibnamefont{Theilmann}},
  \bibinfo{author}{\bibfnamefont{R.}~\bibnamefont{Matzdorf}},
  \bibinfo{author}{\bibfnamefont{G.}~\bibnamefont{Meister}}, \bibnamefont{and}
  \bibinfo{author}{\bibfnamefont{A.}~\bibnamefont{Goldmann}},
  \bibinfo{journal}{Phys. Rev. B} \textbf{\bibinfo{volume}{56}},
  \bibinfo{pages}{3632} (\bibinfo{year}{1997}).

\bibitem[{\citenamefont{Theilmann et~al.}(1999)\citenamefont{Theilmann,
  Matzdorf, and Goldmann}}]{Theilmann199933}
\bibinfo{author}{\bibfnamefont{F.}~\bibnamefont{Theilmann}},
  \bibinfo{author}{\bibfnamefont{R.}~\bibnamefont{Matzdorf}}, \bibnamefont{and}
  \bibinfo{author}{\bibfnamefont{A.}~\bibnamefont{Goldmann}},
  \bibinfo{journal}{Surface Science} \textbf{\bibinfo{volume}{420}},
  \bibinfo{pages}{33 } (\bibinfo{year}{1999}).

\bibitem[{\citenamefont{Fauster et~al.}(2000)\citenamefont{Fauster, Reu\ss{},
  Shumay, Weinelt, Theilmann, and Goldmann}}]{PhysRevB.61.16168}
\bibinfo{author}{\bibfnamefont{T.}~\bibnamefont{Fauster}},
  \bibinfo{author}{\bibfnamefont{C.}~\bibnamefont{Reu\ss{}}},
  \bibinfo{author}{\bibfnamefont{I.~L.} \bibnamefont{Shumay}},
  \bibinfo{author}{\bibfnamefont{M.}~\bibnamefont{Weinelt}},
  \bibinfo{author}{\bibfnamefont{F.}~\bibnamefont{Theilmann}},
  \bibnamefont{and} \bibinfo{author}{\bibfnamefont{A.}~\bibnamefont{Goldmann}},
  \bibinfo{journal}{Phys. Rev. B} \textbf{\bibinfo{volume}{61}},
  \bibinfo{pages}{16168} (\bibinfo{year}{2000}).

\bibitem[{\citenamefont{Quinn}(1962)}]{PhysRev.126.1453}
\bibinfo{author}{\bibfnamefont{J.~J.} \bibnamefont{Quinn}},
  \bibinfo{journal}{Phys. Rev.} \textbf{\bibinfo{volume}{126}},
  \bibinfo{pages}{1453} (\bibinfo{year}{1962}).

\bibitem[{\citenamefont{Hedin and Lundqvist}(1969)}]{GWA}
\bibinfo{author}{\bibfnamefont{L.}~\bibnamefont{Hedin}} \bibnamefont{and}
  \bibinfo{author}{\bibfnamefont{S.}~\bibnamefont{Lundqvist}},
  \bibinfo{journal}{Solid State Physics}  (\bibinfo{year}{1969}).

\bibitem[{\citenamefont{Chulkov et~al.}(1998)\citenamefont{Chulkov,
  Sarr\'\i{}a, Silkin, Pitarke, and Echenique}}]{PhysRevLett.80.4947}
\bibinfo{author}{\bibfnamefont{E.~V.} \bibnamefont{Chulkov}},
  \bibinfo{author}{\bibfnamefont{I.}~\bibnamefont{Sarr\'\i{}a}},
  \bibinfo{author}{\bibfnamefont{V.~M.} \bibnamefont{Silkin}},
  \bibinfo{author}{\bibfnamefont{J.~M.} \bibnamefont{Pitarke}},
  \bibnamefont{and} \bibinfo{author}{\bibfnamefont{P.~M.}
  \bibnamefont{Echenique}}, \bibinfo{journal}{Phys. Rev. Lett.}
  \textbf{\bibinfo{volume}{80}}, \bibinfo{pages}{4947} (\bibinfo{year}{1998}).

\bibitem[{\citenamefont{Eiguren et~al.}(2002)\citenamefont{Eiguren, Hellsing,
  Reinert, Nicolay, Chulkov, Silkin, H\"ufner, and
  Echenique}}]{PhysRevLett.88.066805}
\bibinfo{author}{\bibfnamefont{A.}~\bibnamefont{Eiguren}},
  \bibinfo{author}{\bibfnamefont{B.}~\bibnamefont{Hellsing}},
  \bibinfo{author}{\bibfnamefont{F.}~\bibnamefont{Reinert}},
  \bibinfo{author}{\bibfnamefont{G.}~\bibnamefont{Nicolay}},
  \bibinfo{author}{\bibfnamefont{E.~V.} \bibnamefont{Chulkov}},
  \bibinfo{author}{\bibfnamefont{V.~M.} \bibnamefont{Silkin}},
  \bibinfo{author}{\bibfnamefont{S.}~\bibnamefont{H\"ufner}}, \bibnamefont{and}
  \bibinfo{author}{\bibfnamefont{P.~M.} \bibnamefont{Echenique}},
  \bibinfo{journal}{Phys. Rev. Lett.} \textbf{\bibinfo{volume}{88}},
  \bibinfo{pages}{066805} (\bibinfo{year}{2002}).

\bibitem[{\citenamefont{Eiguren et~al.}(2003)\citenamefont{Eiguren, Hellsing,
  Chulkov, and Echenique}}]{PhysRevB.67.235423}
\bibinfo{author}{\bibfnamefont{A.}~\bibnamefont{Eiguren}},
  \bibinfo{author}{\bibfnamefont{B.}~\bibnamefont{Hellsing}},
  \bibinfo{author}{\bibfnamefont{E.~V.} \bibnamefont{Chulkov}},
  \bibnamefont{and} \bibinfo{author}{\bibfnamefont{P.~M.}
  \bibnamefont{Echenique}}, \bibinfo{journal}{Phys. Rev. B}
  \textbf{\bibinfo{volume}{67}}, \bibinfo{pages}{235423}
  (\bibinfo{year}{2003}).

\bibitem[{\citenamefont{Wiesenmayer et~al.}(2008)\citenamefont{Wiesenmayer,
  Bauer, Mathias, Wessendorf, Chulkov, Silkin, Borisov, Gauyacq, Echenique, and
  Aeschlimann}}]{PhysRevB.78.245410}
\bibinfo{author}{\bibfnamefont{M.}~\bibnamefont{Wiesenmayer}},
  \bibinfo{author}{\bibfnamefont{M.}~\bibnamefont{Bauer}},
  \bibinfo{author}{\bibfnamefont{S.}~\bibnamefont{Mathias}},
  \bibinfo{author}{\bibfnamefont{M.}~\bibnamefont{Wessendorf}},
  \bibinfo{author}{\bibfnamefont{E.~V.} \bibnamefont{Chulkov}},
  \bibinfo{author}{\bibfnamefont{V.~M.} \bibnamefont{Silkin}},
  \bibinfo{author}{\bibfnamefont{A.~G.} \bibnamefont{Borisov}},
  \bibinfo{author}{\bibfnamefont{J.-P.} \bibnamefont{Gauyacq}},
  \bibinfo{author}{\bibfnamefont{P.~M.} \bibnamefont{Echenique}},
  \bibnamefont{and}
  \bibinfo{author}{\bibfnamefont{M.}~\bibnamefont{Aeschlimann}},
  \bibinfo{journal}{Phys. Rev. B} \textbf{\bibinfo{volume}{78}},
  \bibinfo{pages}{245410} (\bibinfo{year}{2008}).

\bibitem[{\citenamefont{Graefenstein et~al.}(1988)\citenamefont{Graefenstein,
  Mertig, and Zeller}}]{Graefenstein88}
\bibinfo{author}{\bibfnamefont{J.}~\bibnamefont{Graefenstein}},
  \bibinfo{author}{\bibfnamefont{I.}~\bibnamefont{Mertig}}, \bibnamefont{and}
  \bibinfo{author}{\bibfnamefont{R.}~\bibnamefont{Zeller}},
  \bibinfo{journal}{J. Phys. F: Met. Phys.} \textbf{\bibinfo{volume}{18}},
  \bibinfo{pages}{731} (\bibinfo{year}{1988}).

\bibitem[{\citenamefont{Mertig et~al.}(1982)\citenamefont{Mertig, Mrosan, and
  Sch\"opke}}]{Mertig82}
\bibinfo{author}{\bibfnamefont{I.}~\bibnamefont{Mertig}},
  \bibinfo{author}{\bibfnamefont{E.}~\bibnamefont{Mrosan}}, \bibnamefont{and}
  \bibinfo{author}{\bibfnamefont{R.}~\bibnamefont{Sch\"opke}},
  \bibinfo{journal}{J. Phys. F: Met. Phys.} \textbf{\bibinfo{volume}{12}},
  \bibinfo{pages}{1689} (\bibinfo{year}{1982}).

\bibitem[{\citenamefont{Hewson}(2003)}]{hewson}
\bibinfo{author}{\bibfnamefont{A.~C.} \bibnamefont{Hewson}},
  \emph{\bibinfo{title}{The Kondo problem to heavy fermions}}
  (\bibinfo{publisher}{Cambridge Univ. Press}, \bibinfo{year}{2003}).

\bibitem[{\citenamefont{Papanikolaou et~al.}(1994)\citenamefont{Papanikolaou,
  Stefanou, and Papastaikoudis}}]{Papanikolaou94}
\bibinfo{author}{\bibfnamefont{N.}~\bibnamefont{Papanikolaou}},
  \bibinfo{author}{\bibfnamefont{N.}~\bibnamefont{Stefanou}}, \bibnamefont{and}
  \bibinfo{author}{\bibfnamefont{C.}~\bibnamefont{Papastaikoudis}},
  \bibinfo{journal}{Phys. Rev. B} \textbf{\bibinfo{volume}{49}},
  \bibinfo{pages}{16117} (\bibinfo{year}{1994}), ISSN
  \bibinfo{issn}{0163-1829}.

\bibitem[{\citenamefont{Mavropoulos et~al.}(1995)\citenamefont{Mavropoulos,
  Papanikolaou, and Stefanou}}]{Mavropoulos95}
\bibinfo{author}{\bibfnamefont{P.}~\bibnamefont{Mavropoulos}},
  \bibinfo{author}{\bibfnamefont{N.}~\bibnamefont{Papanikolaou}},
  \bibnamefont{and} \bibinfo{author}{\bibfnamefont{N.}~\bibnamefont{Stefanou}},
  \bibinfo{journal}{J. Phys.: Condens. Matter} \textbf{\bibinfo{volume}{7}},
  \bibinfo{pages}{4665} (\bibinfo{year}{1995}), ISSN \bibinfo{issn}{0953-8984}.

\bibitem[{\citenamefont{Mertig}(1999)}]{Mertig99}
\bibinfo{author}{\bibfnamefont{I.}~\bibnamefont{Mertig}},
  \bibinfo{journal}{Reports on Progress in Physics}
  \textbf{\bibinfo{volume}{62}}, \bibinfo{pages}{237} (\bibinfo{year}{1999}).

\bibitem[{\citenamefont{Papanikolaou et~al.}(2002)\citenamefont{Papanikolaou,
  Zeller, and Dederichs}}]{Papanikolaou02}
\bibinfo{author}{\bibfnamefont{N.}~\bibnamefont{Papanikolaou}},
  \bibinfo{author}{\bibfnamefont{R.}~\bibnamefont{Zeller}}, \bibnamefont{and}
  \bibinfo{author}{\bibfnamefont{P.~H.} \bibnamefont{Dederichs}},
  \bibinfo{journal}{Journal of Physics: Condensed Matter}
  \textbf{\bibinfo{volume}{14}}, \bibinfo{pages}{2799} (\bibinfo{year}{2002}).

\bibitem[{\citenamefont{Ebert et~al.}(2011)\citenamefont{Ebert, K\"odderitzsch,
  and Minar}}]{Ebert11}
\bibinfo{author}{\bibfnamefont{H.}~\bibnamefont{Ebert}},
  \bibinfo{author}{\bibfnamefont{D.}~\bibnamefont{K\"odderitzsch}},
  \bibnamefont{and} \bibinfo{author}{\bibfnamefont{J.}~\bibnamefont{Minar}},
  \bibinfo{journal}{Rep. Prog. Phys.} \textbf{\bibinfo{volume}{74}},
  \bibinfo{pages}{096501} (\bibinfo{year}{2011}).

\bibitem[{\citenamefont{Wildberger et~al.}(1997)\citenamefont{Wildberger,
  Zeller, and Dederichs}}]{PhysRevB.55.10074}
\bibinfo{author}{\bibfnamefont{K.}~\bibnamefont{Wildberger}},
  \bibinfo{author}{\bibfnamefont{R.}~\bibnamefont{Zeller}}, \bibnamefont{and}
  \bibinfo{author}{\bibfnamefont{P.~H.} \bibnamefont{Dederichs}},
  \bibinfo{journal}{Phys. Rev. B} \textbf{\bibinfo{volume}{55}},
  \bibinfo{pages}{10074} (\bibinfo{year}{1997}).

\bibitem[{\citenamefont{Zeller et~al.}(1995)\citenamefont{Zeller, Dederichs,
  \'Ujfalussy, Szunyogh, and Weinberger}}]{PhysRevB.52.8807}
\bibinfo{author}{\bibfnamefont{R.}~\bibnamefont{Zeller}},
  \bibinfo{author}{\bibfnamefont{P.~H.} \bibnamefont{Dederichs}},
  \bibinfo{author}{\bibfnamefont{B.}~\bibnamefont{\'Ujfalussy}},
  \bibinfo{author}{\bibfnamefont{L.}~\bibnamefont{Szunyogh}}, \bibnamefont{and}
  \bibinfo{author}{\bibfnamefont{P.}~\bibnamefont{Weinberger}},
  \bibinfo{journal}{Phys. Rev. B} \textbf{\bibinfo{volume}{52}},
  \bibinfo{pages}{8807} (\bibinfo{year}{1995}).

\bibitem[{\citenamefont{Vosko et~al.}(1980)\citenamefont{Vosko, Wilk, and
  Nusair}}]{Vosko_Wilk_Nusair}
\bibinfo{author}{\bibfnamefont{S.}~\bibnamefont{Vosko}},
  \bibinfo{author}{\bibfnamefont{L.}~\bibnamefont{Wilk}}, \bibnamefont{and}
  \bibinfo{author}{\bibfnamefont{M.}~\bibnamefont{Nusair}},
  \bibinfo{journal}{Canadian Journal of Physics} \textbf{\bibinfo{volume}{58}},
  \bibinfo{pages}{1200} (\bibinfo{year}{1980}).

\bibitem[{\citenamefont{Shockley}(1939)}]{PhysRev.56.317}
\bibinfo{author}{\bibfnamefont{W.}~\bibnamefont{Shockley}},
  \bibinfo{journal}{Phys. Rev.} \textbf{\bibinfo{volume}{56}},
  \bibinfo{pages}{317} (\bibinfo{year}{1939}).

\bibitem[{\citenamefont{Fedorov et~al.}(2008)\citenamefont{Fedorov, Zahn,
  Gradhand, and Mertig}}]{Spinflip_PhysRevB_2008}
\bibinfo{author}{\bibfnamefont{D.~V.} \bibnamefont{Fedorov}},
  \bibinfo{author}{\bibfnamefont{P.}~\bibnamefont{Zahn}},
  \bibinfo{author}{\bibfnamefont{M.}~\bibnamefont{Gradhand}}, \bibnamefont{and}
  \bibinfo{author}{\bibfnamefont{I.}~\bibnamefont{Mertig}},
  \bibinfo{journal}{Phys. Rev. B} \textbf{\bibinfo{volume}{77}},
  \bibinfo{pages}{092406} (\bibinfo{year}{2008}).

\bibitem[{\citenamefont{Linde}(1931)}]{Linde_1931}
\bibinfo{author}{\bibfnamefont{J.~O.} \bibnamefont{Linde}},
  \bibinfo{journal}{Annalen der Physik} \textbf{\bibinfo{volume}{402}},
  \bibinfo{pages}{52} (\bibinfo{year}{1931}).

\bibitem[{\citenamefont{Linde}(1932)}]{Linde_1932}
\bibinfo{author}{\bibfnamefont{J.~O.} \bibnamefont{Linde}},
  \bibinfo{journal}{Annalen der Physik} \textbf{\bibinfo{volume}{406}},
  \bibinfo{pages}{353} (\bibinfo{year}{1932}).

\bibitem[{per(1995)}]{periodic_table}
\emph{\bibinfo{title}{Periodic table of elements}}
  (\bibinfo{publisher}{Sargent-Welch Scientific}, \bibinfo{year}{1995}).

\bibitem{footnote2}
In photoemission spectroscopy, the surface state of
  silver is found to be ~63 meV below the Fermi energy, while in Cu
  and Au the surface states are much deeper in energy (435 meV and 484
  meV below $E_{\mathrm{F}}$).\cite{PhysRevB.63.115415}


\bibitem[{\citenamefont{Podloucky et~al.}(1980)\citenamefont{Podloucky, Zeller,
  and Dederichs}}]{PhysRevB.22.5777}
\bibinfo{author}{\bibfnamefont{R.}~\bibnamefont{Podloucky}},
  \bibinfo{author}{\bibfnamefont{R.}~\bibnamefont{Zeller}}, \bibnamefont{and}
  \bibinfo{author}{\bibfnamefont{P.~H.} \bibnamefont{Dederichs}},
  \bibinfo{journal}{Phys. Rev. B} \textbf{\bibinfo{volume}{22}},
  \bibinfo{pages}{5777} (\bibinfo{year}{1980}).

\bibitem[{\citenamefont{Grüner}(1974)}]{Gruner74}
\bibinfo{author}{\bibfnamefont{G.}~\bibnamefont{Grüner}},
  \bibinfo{journal}{Adv. Phys.} \textbf{\bibinfo{volume}{23}},
  \bibinfo{pages}{941} (\bibinfo{year}{1974}).

\bibitem[{\citenamefont{Cohen and Slichter}(1978)}]{Cohen78}
\bibinfo{author}{\bibfnamefont{J.}~\bibnamefont{Cohen}} \bibnamefont{and}
  \bibinfo{author}{\bibfnamefont{C.}~\bibnamefont{Slichter}},
  \bibinfo{journal}{J. Appl. Phys.} \textbf{\bibinfo{volume}{49}},
  \bibinfo{pages}{1537} (\bibinfo{year}{1978}).

\bibitem[{\citenamefont{Coleridge}(1985)}]{Coleridge85}
\bibinfo{author}{\bibfnamefont{P.}~\bibnamefont{Coleridge}},
  \bibinfo{journal}{J. Phys. F.: Met. Phys.} \textbf{\bibinfo{volume}{15}},
  \bibinfo{pages}{1727} (\bibinfo{year}{1985}).

\bibitem[{\citenamefont{Matzdorf}(2000)}]{Matzdorf2000151}
\bibinfo{author}{\bibfnamefont{R.}~\bibnamefont{Matzdorf}},
  \bibinfo{journal}{Chemical Physics} \textbf{\bibinfo{volume}{251}},
  \bibinfo{pages}{151 } (\bibinfo{year}{2000}).

\end{thebibliography}
\end{document}